\documentclass[a4paper,11pt]{article}
\usepackage{jheppub} 
\usepackage{lineno}
\usepackage{siunitx}
\usepackage{subcaption}
\usepackage{mathtools}
\usepackage{amsmath}

\newcommand{\specialcell}[2][c]{%
  \begin{tabular}[#1]{@{}c@{}}#2\end{tabular}}
\newcommand{\abs}[1]{\left\vert {#1} \right\vert}

\arxivnumber{2311.18509} 

\title{Search for Hidden Neutrinos at the European Spallation Source: the SHiNESS experiment}

\collaboration{\includegraphics[height=17mm]{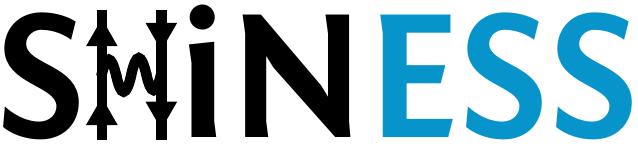}\\[6pt]}

\author[a,1]{Stefano Roberto Soleti,\note{Corresponding author.}}
\author[b]{Pilar Coloma,}
\author[a]{Juan Jose Gómez Cadenas,}
\author[c]{Anatael Cabrera}
\affiliation[a]{Donostia International Physics Center\\
Manuel Lardizabal Ibilbidea, 4, 20018 Donostia, Spain}
\affiliation[b]{Instituto de Física Teórica (IFT-CFTMAT), CSIC-UAM, Calle de Nicolás Cabrera 13–15, Campus de Cantoblanco, E-28049 Madrid, Spain}
\affiliation[c]{2IJCLab, Université Paris-Saclay, CNRS/IN2P3, 91405 Orsay, France}
\preprint{IFT-UAM/CSIC-23-150}

\emailAdd{roberto.soleti@dipc.org}

\abstract{
The upcoming European Spallation Source (ESS) will soon provide the most intense neutrino source in the world. We propose the Search for Hidden Neutrinos at the ESS (SHiNESS) experiment, highlighting its unique opportunities to search for the existence of sterile neutrinos across a wide range of scales: anomalous oscillations at short baselines; non-unitarity mixing in the active neutrino sector; or an excess of events with multiple leptons in the final state, produced in the decay of heavy neutrinos. The baseline design of the detector comprises an active volume filled with 42~ton of liquid scintillator, located 25~m far from the ESS beam target. We show that SHiNESS will be able to considerably improve current global limits for the three cases outlined above. Although in this work we focus on new physics in the neutrino sector, the proposed setup may also be used to search for signals from weakly interacting particles in a broader context. }
\begin{document}
\maketitle
\flushbottom

\section{\label{sec:intro}Introduction}
A plethora of neutrino experiments carried out in the last few decades have confirmed existence of neutrino oscillations, which implies nonzero neutrino masses~\cite{Super-Kamiokande:1998kpq, SNO:2002tuh, KamLAND:2004mhv}. Neutrino mixing can be described by a unitary matrix that parametrizes the change of basis between the flavor states and mass eigenstates. After the measurement of the last mixing angle $\theta_{13}$~\cite{DayaBay:2012fng}, our knowledge of the parameters that govern neutrino mixing in three families is almost complete, with the notable exception of the CP-violating phase and the neutrino mass ordering, which are still unknown. In the meantime, the neutrino research program has entered a \emph{precision} era~\cite{Esteban:2020cvm, T2K:2021xwb, NOvA:2021nfi} in the realm of neutrino oscillations, which will be further improved with the upcoming experiments DUNE~\cite{DUNE:2015lol}, JUNO~\cite{JUNO:2020bcl} and Hyper-Kamiokande~\cite{Abe:2011ts}. 

However, our understanding of the neutrino sector is still limited. Although the observation of neutrino oscillations implies that at least two neutrinos are massive, the values of their masses remain unknown. Moreover, the Standard Model (SM) of particle physics does not provide a mechanism to give neutrinos a mass, which necessarily requires to enlarge the SM particle content. Even in the simplest scenario where right-handed neutrinos are added to the SM in order to generate neutrino masses, new questions arise regarding the specific mechanism at play, the Majorana/Dirac nature of neutrinos, and the scale of new physics associated to the new particles introduced. In fact, neutrinos offer a window to explore multiple effects stemming from new physics beyond the SM (BSM) across a wide range of scales. 

The existence of additional neutrino mass eigenstates would have different phenomenological consequences, depending on the scale associated to the new particles. In fact, leaving theoretical arguments aside, the existence of new neutrino mass eigenstates (and the scale of new physics associated to their mass) can only be probed experimentally. While precision measurements at collider experiments have confirmed that the number of light neutrinos that participate in the weak interactions is three~\cite{ALEPH:2005ab}, a number of neutrino experiments have produced results not in full agreement with this landscape. These somehow controversial \emph{short-baseline} anomalies include a significant excess of electron (anti)neutrinos in a mainly muon (anti)neutrino beam, as observed by the LSND~\cite{LSND:2001aii} and MiniBooNE~\cite{MiniBooNE:2020pnu} experiments (the \emph{LSND/MiniBooNE} anomalies, each of them with a significance above $4\sigma$); and an evidence of electron neutrino disappearance observed by several experiments employing a gallium detector exposed to a radioactive source~\cite{Abdurashitov:2005tb, GALLEX:1994rym, Barinov:2021asz} (the \emph{gallium anomaly}~\cite{Laveder:2007zz,Acero:2007su,Giunti:2010zu}, 
with a significance above $5\sigma$~\cite{Barinov:2022wfh, Barinov:2021mjj, Berryman:2021yan, Elliott:2023cvh, Brdar:2023cms}). 
The most naive explanation for these anomalies is the existence of a new neutrino with a mass squared splitting of around 1~eV$^2$, called \emph{light sterile neutrinos}. 
However, the experimental landscape is not fully consistent~\cite{Boser:2019rta}, given the absence of an excess in the KARMEN~\cite{KARMEN:2002zcm} and MicroBooNE~\cite{MicroBooNE:2021tya} experiments, the significant tension between appearance and disappearance measurements~\cite{MiniBooNE:2012meu, IceCube:2016rnb, Dentler:2018sju, Diaz:2019fwt}, and that the so-called \emph{reactor antineutrino anomaly} has recently vanished in light of new reactor flux reevaluations~\cite{Berryman:2019hme, Estienne:2019ujo, Giunti:2021kab}. Thus, the community still awaits new experimental data that could shed some light on the origin of the reported anomalies.

Sterile neutrinos at higher scales may not lead to a new oscillation frequency but could yield interesting phenomenological signals elsewhere. To begin with, additional neutrinos could induce an apparent violation of the unitarity of the neutrino mixing matrix in the $3\times 3$ active block (which governs oscillations among active neutrinos), potentially even involving additional sources of CP violation~\cite{Langacker:1988up,Bilenky:1992wv,Bergmann:1998rg,Czakon:2001em,Bekman:2002zk,Antusch:2006vwa,Fernandez-Martinez:2007iaa,Escrihuela:2016ube}. While non-unitarity constraints from heavy neutrinos (above the EW scale) are severely constrained from global fits to electroweak precision data~\cite{Blennow:2023mqx,Antusch:2016brq}, neutrinos at lower scales could still lead to observable effects~\cite{Blennow:2016jkn}. For neutrino masses below the pion mass, the effect would be that of an averaged-out oscillation, leading to an overall constant effect on the transition probabilities. On the other hand, a sterile neutrino with a mass in the MeV-GeV range, usually called \emph{heavy neutral lepton} (HNL), may be produced in meson decays through its mixing with the active neutrinos (for recent reviews on HNL phenomenology, see e.g. refs.~\cite{Abdullahi:2022jlv,Antel:2023hkf}). After propagating over some distance, it could decay back into SM particles, leaving a visible signal in neutrino detectors. Typical decay signals for a HNL below the GeV scale that only interacts via mixing involve a combination of light mesons, electrons/muons and neutrinos (which lead to missing energy).

In view of the general features outlined above, an intense, pulsed, source of well-characterized neutrinos would be the ideal tool to probe for any of the new physics signals outlined above. Spallation sources are excellent in this regard: in the proton collisions in the target, an intense flux of pions is produced that in turn yields a high-intensity neutrino flux from their decay at rest (DAR). The upcoming European Spallation Source (ESS) currently under construction in Lund, Sweden, will provide the most intense neutron beam in the world~\cite{Garoby:2017vew} and therefore offers an exceptional opportunity to search for new physics in the neutrino sector. To this end, we propose the Search for Hidden Neutrinos at the European Spallation Source (SHiNESS) experiment, which will be able to: (1) probe the unitarity of the active neutrino mixing matrix with unprecedented sensitivity; (2) set world-leading limits on heavy neutral leptons, and (3) definitely clarify whether the origin of the short-baseline neutrino anomalies are due to an eV-scale sterile neutrino. We note that, while studies exist to use the future ESS Neutrino Super Beam~\cite{Alekou:2022emd} (ESS$\nu$SB) for light sterile neutrino searches~\cite{Blennow:2014fqa, KumarAgarwalla:2019blx, Ghosh:2019zvl}, SHiNESS does not require any upgrade to the beamline and can start acquiring data as soon as the ESS turns on. Moreover, we highlight its complementarity to the current neutrino physics program of the ESS~\cite{Baxter:2019mcx}, which focuses on coherent elastic neutrino-nucleus scattering (CE$\nu$NS) detection to pursue new physics signals: both experiments may take place simultaneously without interfering with any of the main physics goals of the ESS (or with each other).

This document is organized as follows. Section~\ref{sec:setup} describes in detail the experimental setup at the ESS and the detector technology, while section~\ref{sec:pheno} discusses the phenomenological signals expected for the three cases under consideration in this work: zero-distance effects induced by a non-unitarity leptonic mixing matrix, oscillations at short baselines induced by light sterile neutrinos, and decay signals from heavy neutral leptons. Section~\ref{sec:principle} discusses the expected number of events for each case, while section~\ref{sec:bkg} describes the main backgrounds expected. Section~\ref{sec:results} contains the calculated sensitivity of the experiment to the different scenarios considered. A summary of the proposal and a comparison with the current experimental landscape are provided in section~\ref{sec:summary}.

\section{\label{sec:setup}The experimental setup}
\subsection{\label{sec:beam}The beam}
\begin{figure}[htbp]
\centering
\includegraphics[width=0.7\textwidth]{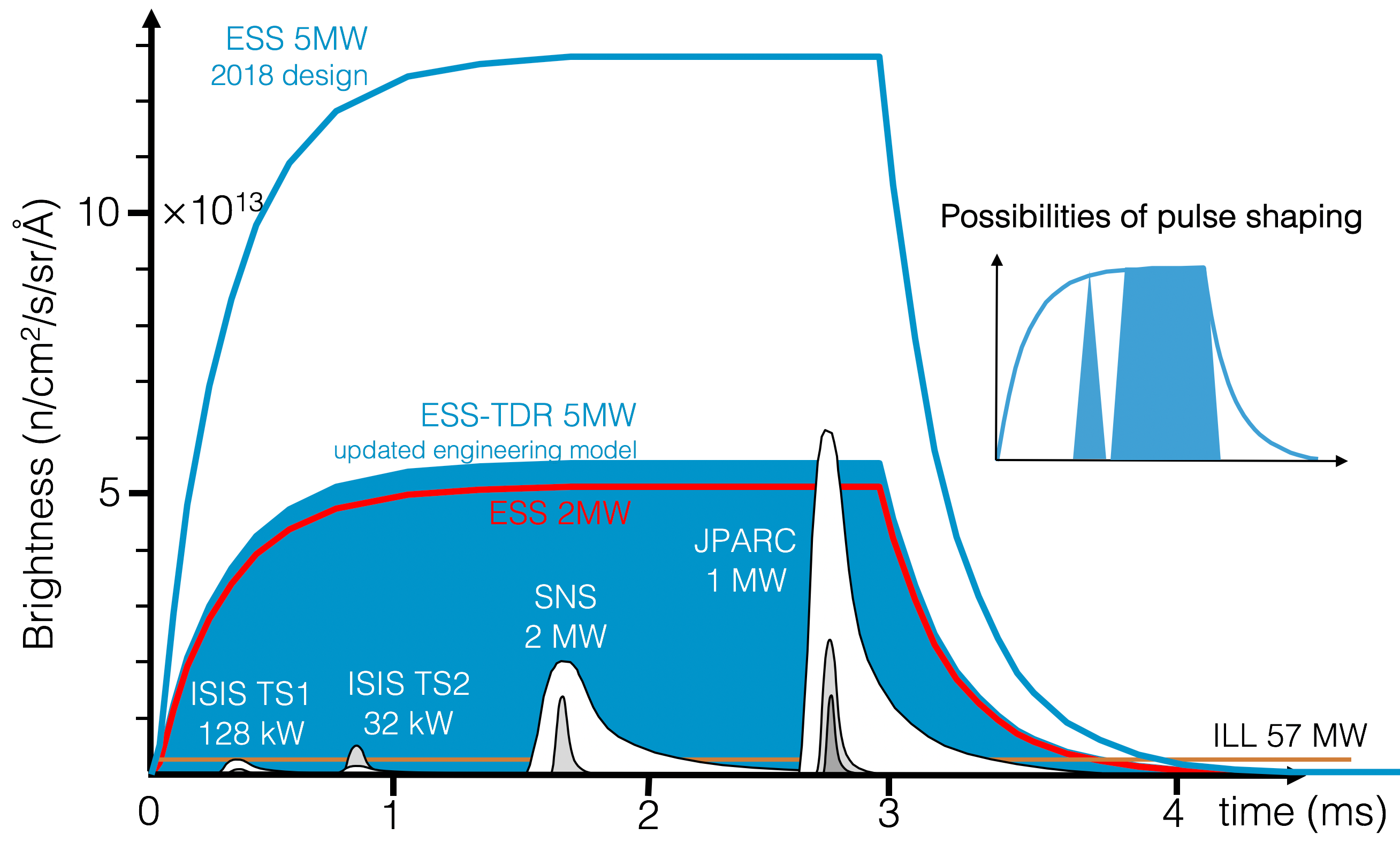}
\caption{\label{fig:essbeam} ESS pulse structure in terms of neutron brightness compared to other facilities. The total number of neutrons is proportional to the area under the curve. Adapted from ref.~\cite{GUTHRIE2017637}.}
\end{figure}

The European Spallation Source is an upcoming facility which will combine a 2 GeV proton linac with an advanced hydrogen moderator, generating the most intense neutron beam in the world~\cite{GUTHRIE2017637}. The proton beam impinges on a rotating tungsten target, producing fast, high-energy neutrons that are slowed down by the moderator. 
The 5 MW design power will deliver a total of $2.8\times10^{23}$ protons-on-target per calendar year, assuming 5000 hours of beam delivery. The proton beam pulse has a rate of 14 Hz with 2.8 ms long spills, giving a duty factor of $4\times10^{-2}$. A comparison of the ESS beam pulse with other spallation sources is shown in figure \ref{fig:essbeam}.

\begin{figure}[htbp]
\centering
\includegraphics[width=0.7\textwidth]{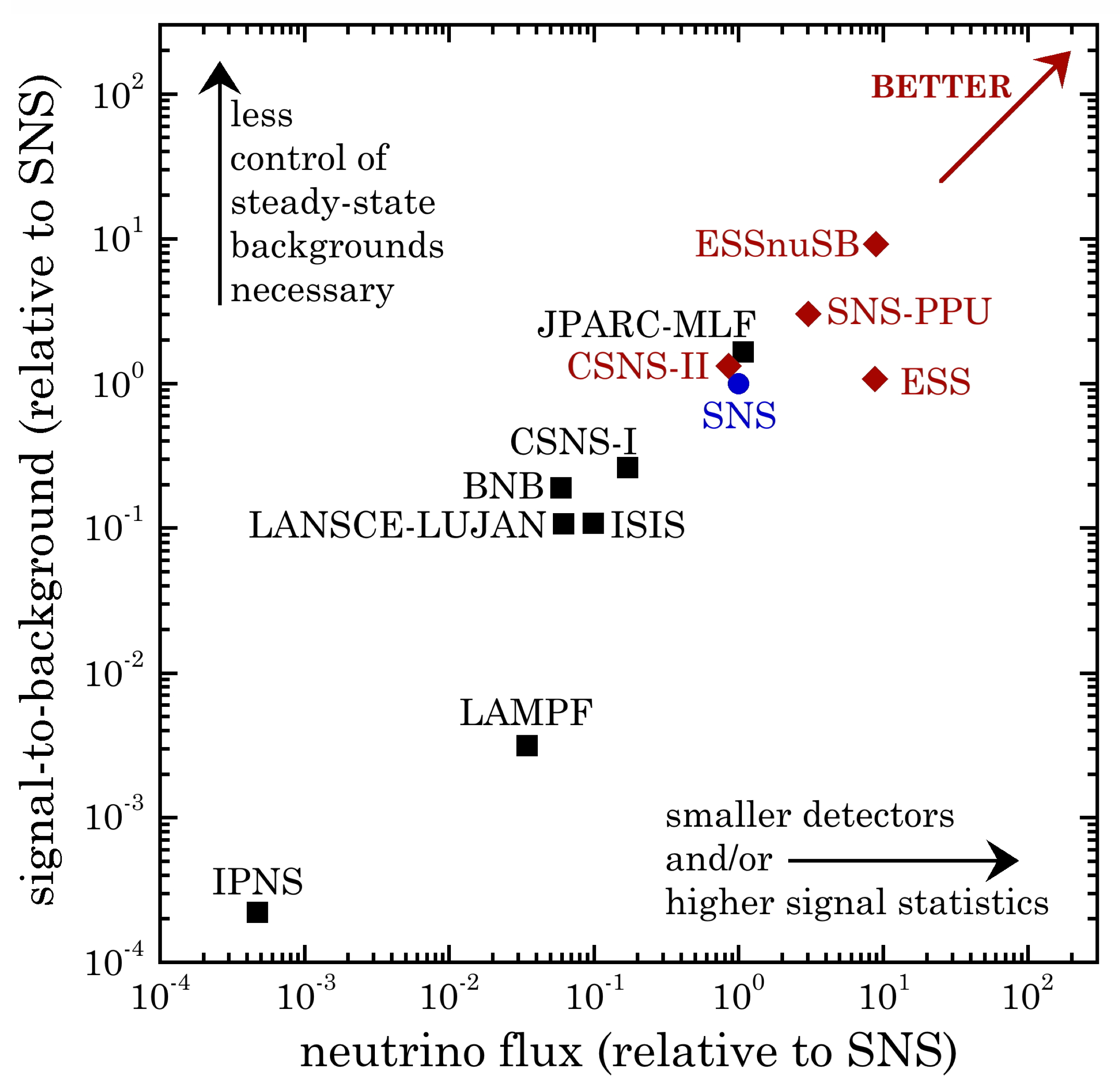}
\caption{\label{fig:comparison} Comparison of past (black squares) and future (red diamonds) spallation sources relative to the SNS (blue circle), in terms of neutrino flux and signal-to-background FOM. Adapted from ref.~\cite{Abele:2022iml}. }
\end{figure}

Compared to the Spallation Neutron Source (SNS) in Oak Ridge, United States and the J-PARC facility in Tokai, Japan, the ESS has a much larger duty factor. This feature, in principle, could increase the background from steady-state processes such as cosmic rays. However, steady-state backgrounds can be accurately characterized during the long anti-coincident periods between beam spills \cite{Baxter:2019mcx}, and the signal-to-background figure of merit (FOM), as defined in ref.~\cite{Abele:2022iml} and reproduced in figure \ref{fig:comparison}, is comparable for the three facilities.

Spallation sources produce both $\pi^-$ and $\pi^+$ in the proton-nucleus collisions in the target. Negative pions are efficiently absorbed by nuclei before they can decay, while positive ones lose energy as they propagate and finally decay at rest through:
\begin{align}
    \pi^+\rightarrow&\mu^+ + \nu_{\mu}\label{eq:dar}\\
                    &\mu^+\rightarrow e^++ \nu_e+\bar{\nu}_{\mu} \nonumber.
\end{align}
The energy spectra produced by these processes are well known~\cite{Burman:2003ek,Baxter:2019mcx}. The $\nu_{\mu}$ flux is monochromatic with $E_{\nu_{\mu}}\approx29.7$~MeV, since the $\pi^+$ decay is a two-body process, while the $\nu_e$ and the $\bar{\nu}_{\mu}$ fluxes exhibit a Michel distribution at energies $E_{\nu_e,\bar{\nu}_{\mu}}<m_{\mu}/2\approx52.8$~MeV. The goal of SHiNESS is to compare the expected amounts of $\pi^+$ DAR neutrino interactions with the measured ones. 

\begin{figure}[htbp]
\centering
\includegraphics[width=0.7\textwidth]{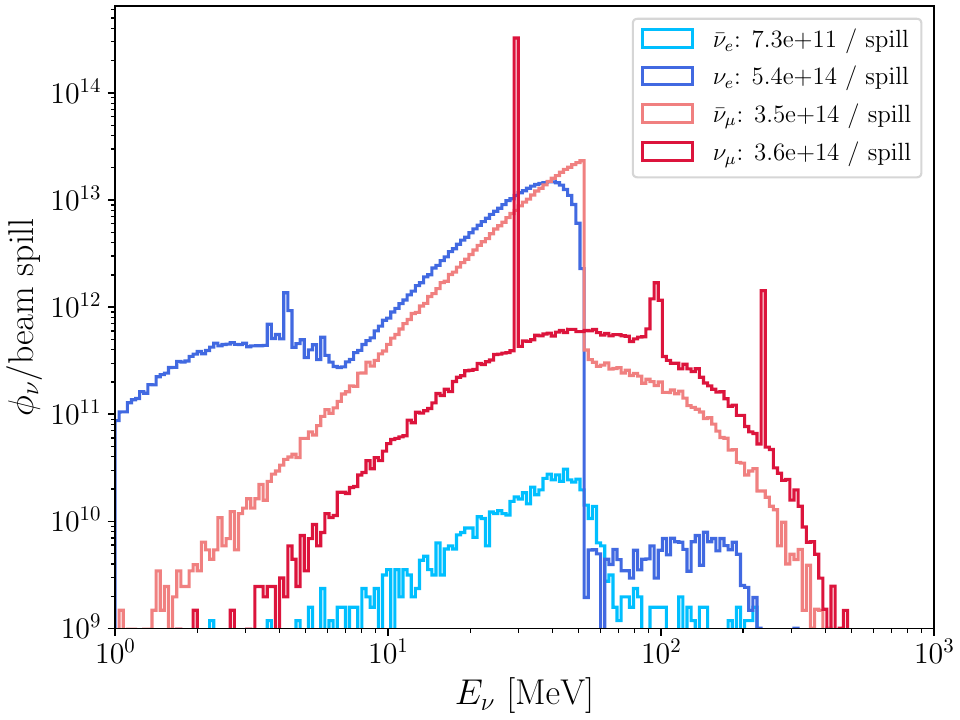}
\caption{\label{fig:nubeam} Flux of each neutrino flavor for an ESS beam spill as a function of the neutrino energy. The simulation has been performed with Geant4 using the BERT-HP physics list.}
\end{figure}

A simplified simulation of the beam was performed for exploratory studies regarding the detection of coherent elastic neutrino-nucleus scattering (CE$\nu$NS) at the ESS. The neutrino yield obtained by this simulation is 0.3 per proton~\cite{Baxter:2019mcx}, resulting in an expected $8.5\times10^{22}$ $\pi^+$ DAR neutrinos per flavor per year. Figure \ref{fig:nubeam} shows the neutrino flux as a function of the energy, obtained with a Geant4~\cite{GEANT4:2002zbu} simulation using the BERT-HP physics list~\cite{Heikkinen:2003sc}. 

Apart from the monoenergetic peak at 29.7~MeV and the two Michel distributions for the $\bar{\nu}_{\mu}$ and the $\nu_e$ components, the spectrum presents several interesting features: \begin{itemize}
    \item muons captured by an atom can decay in orbit, producing $\nu_{\mu}$ with an endpoint energy of around 105 MeV;
    \item kaons produced by the beam impinging on the target can also decay and produce monoenergetic 236 MeV $\nu_{\mu}$ through $K^+\rightarrow\mu^+\nu_{\mu}$;
    \item the small amount of $\mu^-$ that decay before being absorbed can produce $\bar{\nu}_e$ through $\mu^-\rightarrow e^-\bar{\nu}_{e}\nu_{\mu}$. This component represents an irreducible background for the $\bar{\nu}_e$ appearance search (see section \ref{sec:nue}).
\end{itemize}

The COHERENT collaboration has shown that the systematic uncertainty associated to the neutrino flux of the SNS can be constrained at the 10\% level~\cite{COHERENT:2021yvp}. This is also the uncertainty assumed in the proposal for CE$\nu$NS detection at the ESS~\cite{Baxter:2019mcx}. Thus, in this document, we adopt a 10\% systematic uncertainty for the $\nu_e$, $\nu_{\mu}$, and $\bar{\nu}_{\mu}$ components. However, the $\bar{\nu}_e$ component from $\mu^-$ decay has a very poor normalization constraint, since the $\mu^-$ decay and pion production points are not well known. This component can depend on the detail of the simulation and on the physics model adopted, so we assign a systematic uncertainty of 25\%.

Total neutrino flux measurement could be performed through the neutral current (NC) channel of eq.~\eqref{eq:nc} (see section \ref{sec:nc}). The detector could then be used also as a neutrino flux monitor for the proposed CE$\nu$NS experiments at the ESS~\cite{Baxter:2019mcx}. The CE$\nu$NS detection proposal also mentions the possibility to replicate at the ESS a small heavy water detector (1 m$^3$ D$_2$O) proposed to independently measure the neutrino yield of the SNS~\cite{COHERENT:2021xhx}.

\subsection{\label{sec:detector}The detector}
\subsubsection{Location}
\begin{figure}[htbp]
\centering
\includegraphics[width=0.7\textwidth]{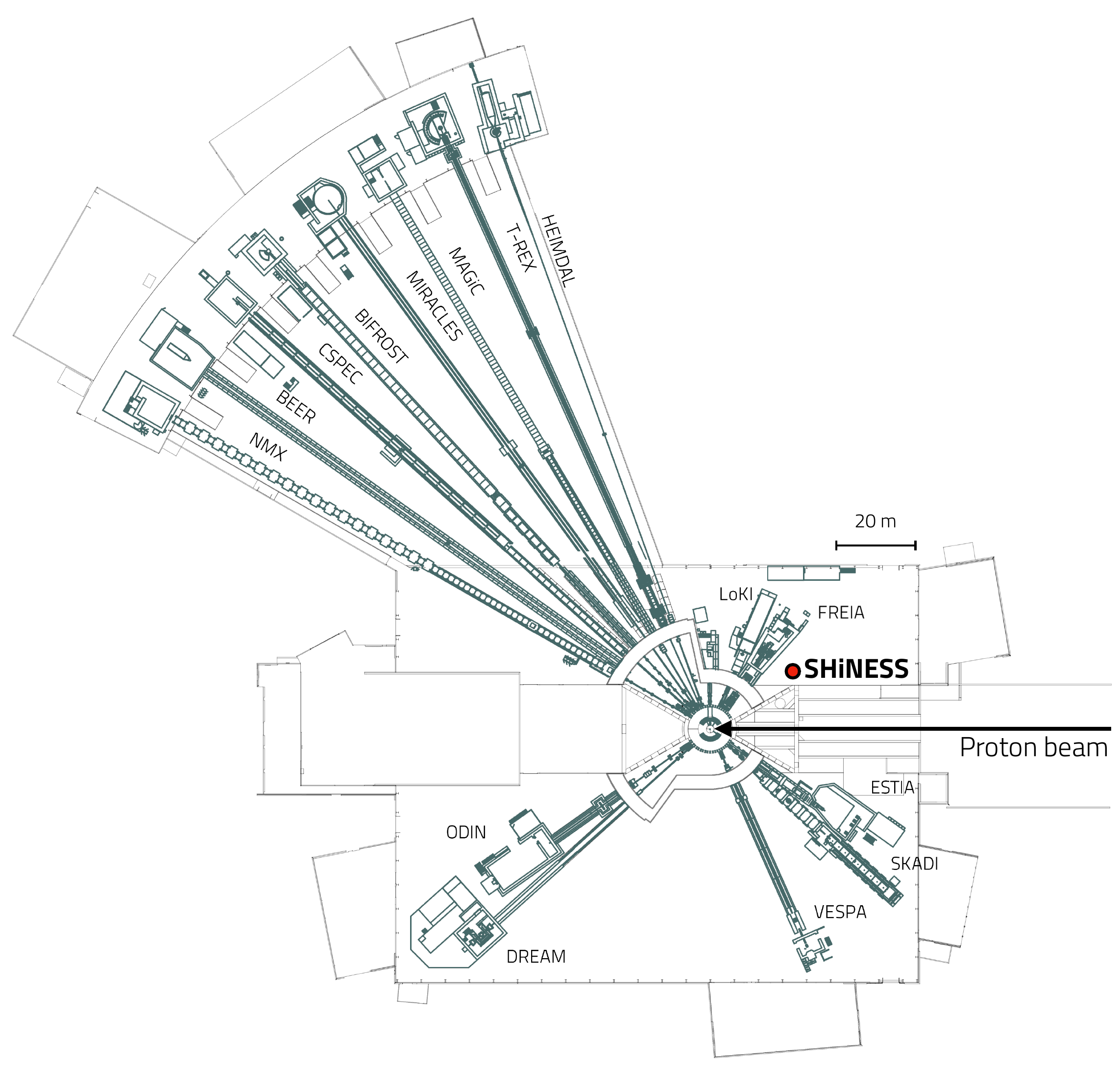}
\caption{\label{fig:plan} Layout of the ESS experimental areas. The proton beam direction is from right to left. The red circle represents the proposed SHiNESS tank, drawn to scale.}
\end{figure}

The D03 experimental hall of the ESS offers unallocated space for a medium-scale experiment. The detector can be placed around 25~m far from the proton beam target, with an angle of approximately $35^{\circ}$ in the backward direction of the beam, in order to suppress forward-boosted particles (see figure \ref{fig:plan}). 

In the short baseline limit of $\Delta m_{21}^2L/E_{\nu} \ll 1$ and $\Delta m_{31}^2L/E_{\nu} \ll 1$, ordinary, three-flavor neutrino oscillations have yet to develop~\cite{Kopp:2013vaa}. Thus, the eventual departure from unitarity will manifest itself through zero-distance effects, as detailed in section \ref{sec:unitarity} and heavy neutral leptons, described in section \ref{sec:hnl}, can be long-lived enough to reach the detector.
In addition, at this distance from the beam target, the $L/E$ of the neutrino oscillation is also well matched to the LSND allowed region in the (3+1) light sterile neutrino model, as shown in section \ref{sec:lightsterile}.

\subsubsection{Technology}

The detection of $\pi^+$DAR neutrino interactions can be efficiently performed with a detector filled with liquid scintillator, which is the solution adopted by the KARMEN~\cite{Zeitnitz:1994kz}, LSND~\cite{LSND:1996jxj}, and JSNS$^2$~\cite{JSNS2:2013jdh} experiments. In this proposal, as a baseline approach, we consider a 5.3~m-high stainless steel cylindrical tank with a radius of 3.3~m. The tank contains an acrylic vessel 4~m high and with a radius of 2~m. The vessel is filled with 42~ton of liquid scintillator loaded with gadolinium. The space between the vessel and the tank inner surface is filled with 31~ton of \emph{unloaded} liquid scintillator, which allows to precisely determine the active volume.

We adopt 2,5-diphenyhloxazole (PPO) as the fluor and linear alkylbenzene C$_6$H$_5$C$_{10-13}$H$_{2(10-13)+1}$ (LAB) as the solvent. This cocktail is largely available, has a high light yield and a high flash point, with a successful track record in neutrino physics~\cite{Beriguete:2014gua, JUNO:2020bcl, SNO:2020fhu, Park:2013nsa}. A possible alternative is represented by a mixture of phenyl-xylylethane (PXE) and dodecane (C$_{12}$H$_{26}$), which was successfully used by the Double Chooz experiment~\cite{DoubleChooz:2006vya, Aberle:2011ar}. The presence of a small concentration of gadolinium (around 0.1\% in mass) greatly increases the neutron absorption probability, given its very high neutron cross section (48800 barn)~\cite{leinweber2006neutron}.
The Gd atom releases a sum of 8~MeV energy in a $\gamma$ cascade after neutron capture~\cite{Tanaka:2019hti}, which has the advantage of being sensibly higher than the energy released upon capture on hydrogen (2.2~MeV). 

\begin{figure}[htbp]
\centering
\includegraphics[width=0.7\columnwidth]{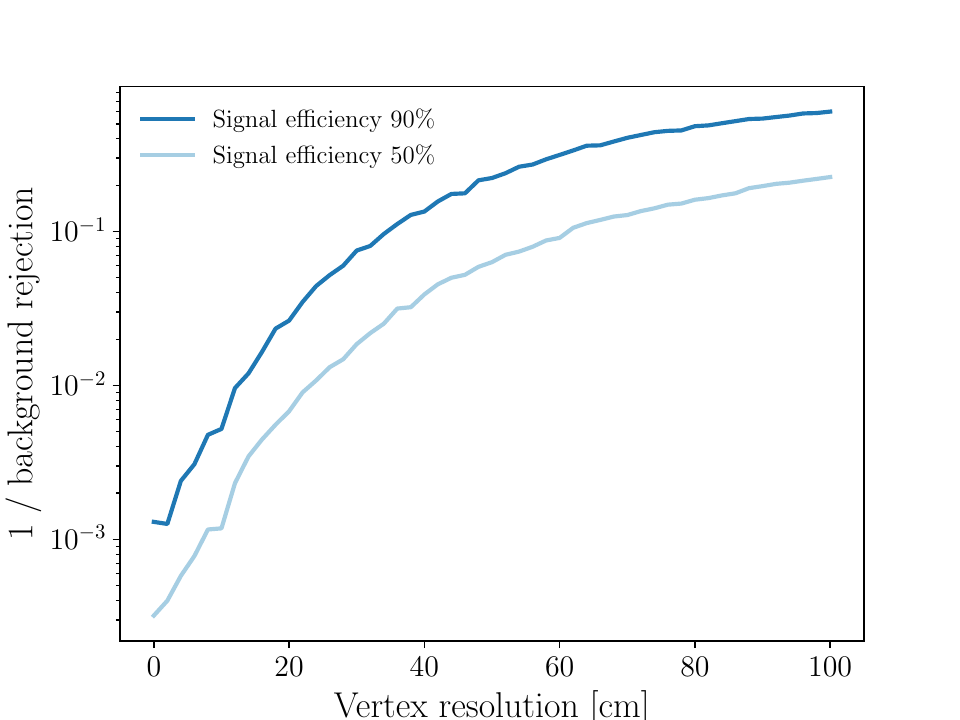}
\caption{\label{fig:pos_res_acc} Background rejection for IBD events as a function of the vertex resolution, assuming a 90\%  (dark blue) and a 50\% (light blue) signal selection efficiency.}
\end{figure}

The design of the SHiNESS detector is driven by the following considerations:
\begin{itemize}
    \item in order to reach competitive sensitivities on the non-unitarity of the neutrino mixing matrix (section \ref{sec:unitarity}) and on the light sterile neutrino searches (section \ref{sec:lightsterile}), accidental backgrounds need to be rejected at a level of at least $10^2$, as shown in section~\ref{sec:acc}.
    \item the detection of HNLs producing a $e^+e^-$ pair (section \ref{sec:hnl}) requires the ability to reconstruct the direction of the two leptons.
\end{itemize}
While signal events for inverse beta decay (IBD) and carbon charged-current interactions produce prompt and delayed signals in close vicinity (see section~\ref{sec:principle}), interactions caused by accidental coincidences will be usually separated by tens of centimeters. Thus, a good vertex resolution is essential to reject this background. Figure~\ref{fig:pos_res_acc} shows the background rejection power as a function of the position resolution. If we require a signal efficiency of 50\%, a vertex resolution of 25~cm corresponds to an accidental background rejection factor of approximately $10^2$. 

In order to satisfy these requirements, the SHiNESS baseline design includes an array of 38 Hamamatsu photomultipliers R12860~\cite{Hamamatsu} with a diameter of 20". 

The detection of Cherenkov photons, in addition to the scintillation light, can improve the vertex resolution and allow to reconstruct the directionality of the particles produced in the interaction, thus enabling HNL searches. 
Separation of Cherenkov and scintillation light can be achieved through timing, directionality, or chromatic separation (or a combination of the three). This is because Cherenkov light is emitted promptly and along the direction of the particle, while scintillation light has a typical emission time constant and is emitted isotropically. Also, the scintillation emission spectrum is relatively narrow, while the Cherenkov radiation follows a power-law distribution given by the Frank-Tamm formula~\cite{Frank:1937fk}. Separation has been demonstrated in LAB and LAB/PPO exploiting both the different time profile~\cite{Li:2015phc, Kaptanoglu:2018sus} and the wavelength spectrum~\cite{Kaptanoglu:2018sus} of the two types of light emission. 

\begin{figure}[htbp]
\centering
\includegraphics[width=0.7\columnwidth]{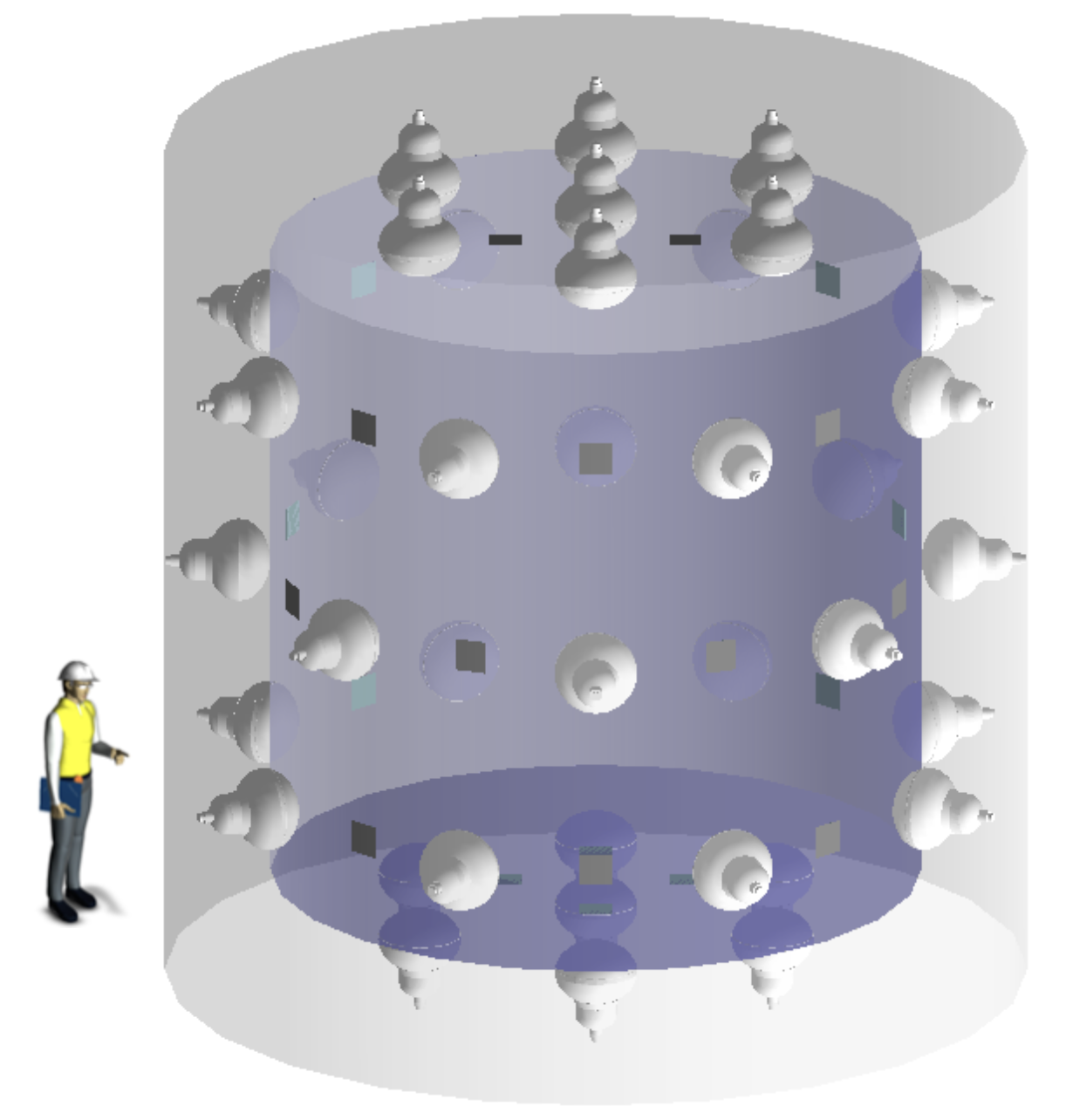}
\caption{\label{fig:detector} Three-dimensional drawing of the SHiNESS detector, which comprises of a cylindrical stainless steel tank (5.3~m high and 5.3~m wide) containing an acrylic vessel (in blue) filled with 42~t of liquid scintillator. The light is collected by 38 Hamamatsu PMTs R12860 (in white) and 32 Incom LAPPDs (the gray squares).}
\end{figure}

In the baseline SHiNESS design, shown in figure \ref{fig:detector}, separation between Cherenkov and scintillation time is enabled by 32 Incom Large Area Picosecond Photodetectors (LAPPDs\textsuperscript{TM})~\cite{Incom} of $20\times20$~cm$^2$. LAPPDs offer single photoelectron time resolutions below 100~ps and sub-cm spatial resolution~\cite{Lyashenko:2019tdj}, allowing to discriminate in time between scintillation and Cherenkov light emission. As a reference, the ANNIE experiment is deploying one LAPPD for every 25 PMTs (whose diameters vary between 8" and 11"), with an expected vertex resolution improvement of a factor of three (from 38~cm to 12~cm) and an expected angular resolution improvement of a factor of two (from 11$^\circ$ to 5$^\circ$) ~\cite{ANNIE:2017nng}. In the case of SHiNESS, 32 LAPPDs allow to detect at least 10 Cherenkov photons in the first 5 ns for 50\% of the inverse beta decay (IBD) events. 

The photosensors, placed uniformly on all sides, provide a photocoverage of 10.1\%. Assuming a light yield of $11,590$ photons/MeV~\cite{GUO201933}, this photocoverage corresponds to an energy resolution of $\sim15\%/\sqrt{E / \mathrm{MeV}}$. The energy resolution is not a critical parameter for the physics searches and the main driver of the photocoverage is the ability to accurately reconstruct the interaction vertex. We stress that the exact number and location of the photosensors can be further optimized once a detailed simulation and an accurate reconstruction chain are available.

\begin{figure}[htbp]
\centering
\includegraphics[width=0.7\columnwidth]{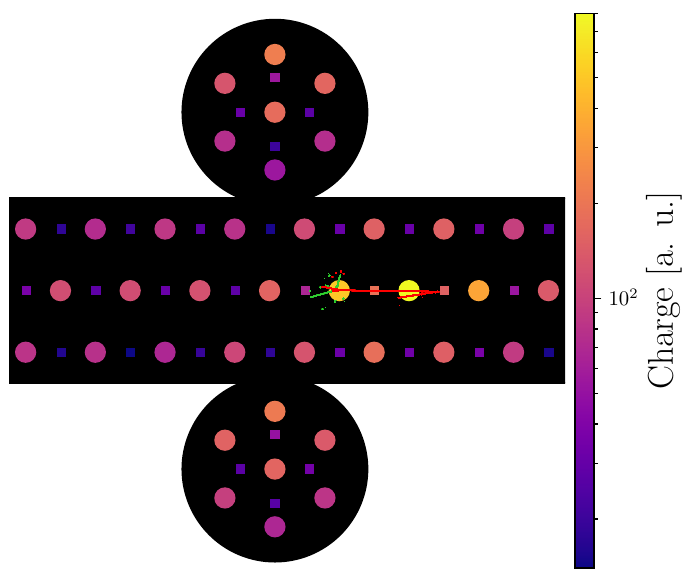}
\caption{\label{fig:evd} Simulated event display of an inverse beta decay in the liquid scintillator tank, showing the charge collected by the photosensors. The circles correspond to Hamamatsu R12860 PMTs and the squares to Incom LAPPDs. The true trajectories correspond to the colored segments (in red the particle tracks generated by the positron, in green the ones generated by the neutron).}
\end{figure}

A simulated event display of an IBD event in the detector (process described in detail in section \ref{sec:ibd}) is shown in figure \ref{fig:evd}. The positron emission and the neutron absorption happen in close vicinity. 

Since the detector will have only a limited overburden, a cosmic-muon veto must be implemented. Two main options are being considered. The first one is represented by an outer layer of liquid scintillator surrounding the main detector and equipped with PMTs. This is the solution adopted by the LSND experiment, which employed also an internal layer of lead shot and achieved a cosmic-ray veto inefficiency $<10^{-5}$ \cite{Conrad:2013mka}. The JSNS$^2$ experiment employs two outer sections of liquid scintillator, with the extra one acting as \emph{gamma-catcher} \cite{JSNS2:2021hyk}. A second option is represented by panels made of plastic scintillator which surround the detector, as implemented by e.g. the KARMEN~\cite{KARMEN:1994xse} and MicroBooNE~\cite{MicroBooNE:2019lta} experiments. 

\subsubsection{Performances}

\begin{figure}[htbp]
\begin{subfigure}[t]{0.49\textwidth}
\includegraphics[width=1\textwidth]{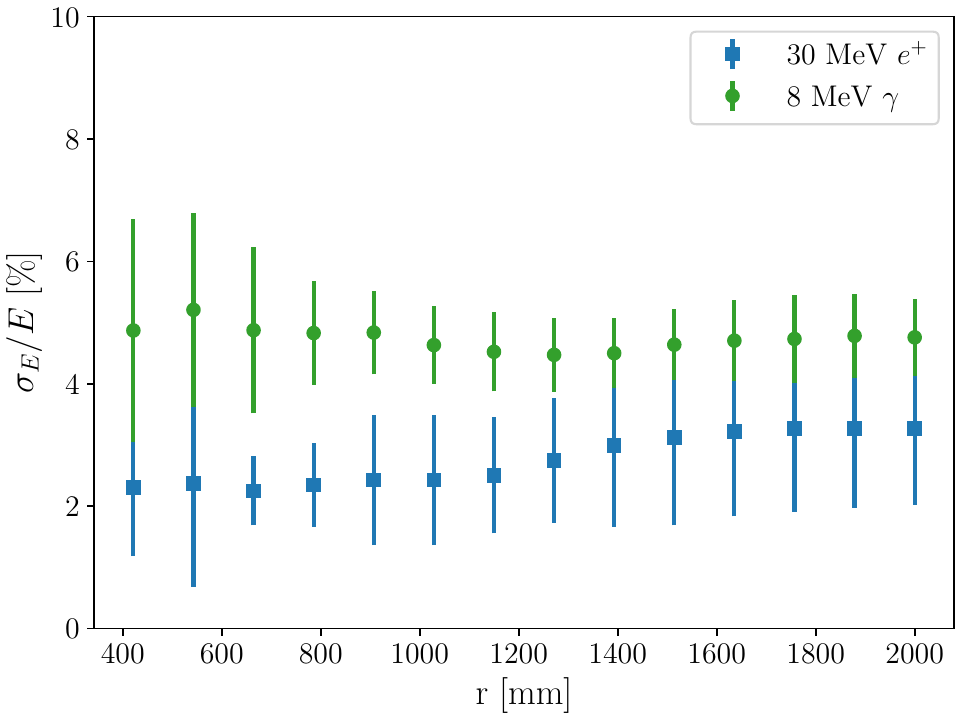}
\caption{\label{fig:resolution} Energy resolution as a function of the distance $r$ from the detector center for a 8 MeV $\gamma$, corresponding to a neutron absorption, and a 30 MeV $e^+$.}
\end{subfigure}
\hfill
\begin{subfigure}[t]{0.49\textwidth}
\includegraphics[width=1\textwidth]{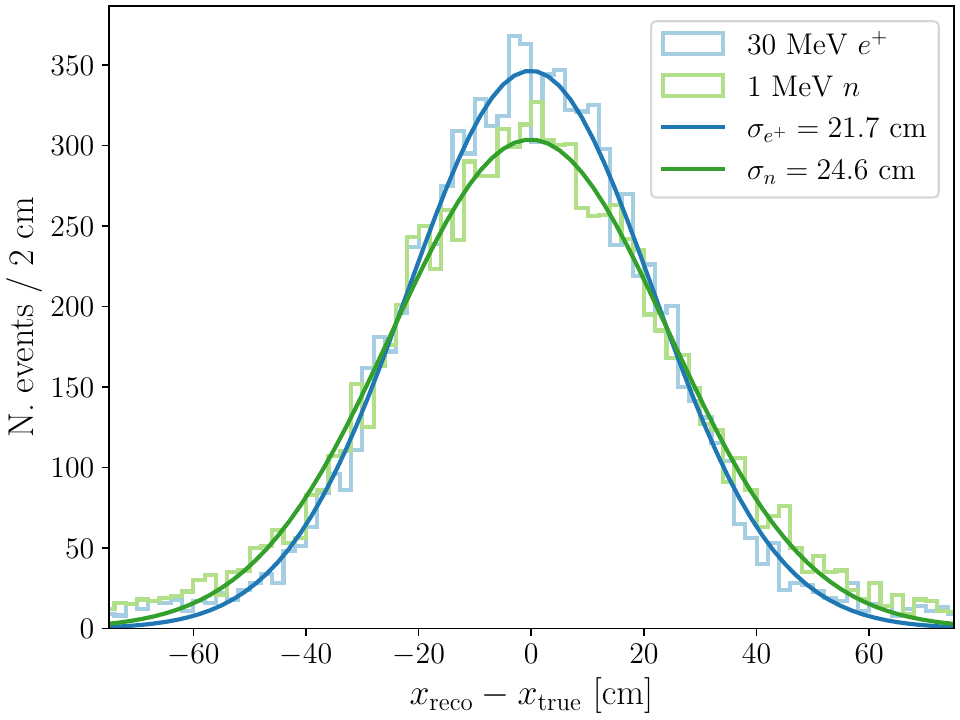}
\caption{\label{fig:position_res} Position resolution in the $x$ axis for a 20~MeV $e^+$ and a 1~MeV neutron. The $x$ coordinate has been reconstructed with a centroid algorithm using the scintillation light in the PMTs.}
\end{subfigure}
\caption{\label{fig:res} Energy (left) and position (right) resolution in the SHiNESS detector for a positron and a neutron.}
\end{figure}

The long attenuation length of the liquid scintillator cocktail (14~m at 430~nm for LAB/PPO \cite{Beriguete:2014gua}), combined with a uniform photosensor coverage, allows achieving an almost uniform energy resolution as a function of the radius, as shown in figure \ref{fig:resolution} for a Geant4 simulation of 30~MeV positrons and 8~MeV Gd-capture $\gamma$.

A naive centroid algorithm using only the light detected by the PMTs gives a vertex resolution of approximately 20~cm (see figure \ref{fig:position_res}). This value can be significantly improved by reconstructing the Cherenkov cone using the first hundreds of picoseconds of the LAPPDs signals, as demonstrated by the ANNIE experiment \cite{ANNIE:2017nng}. 

\begin{figure}[htbp]
\centering
\includegraphics[width=1\columnwidth]{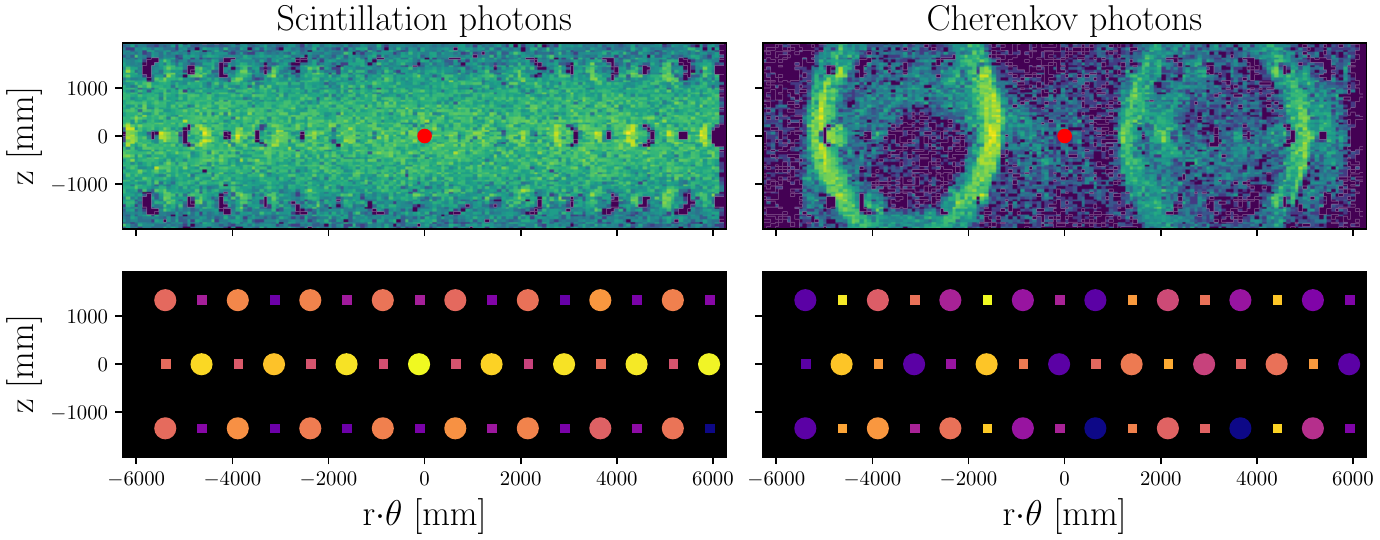}
\caption{\label{fig:cherenkov_event} Spatial distribution of scintillation and Cherenkov photons (top) and the corresponding charge collected by the photosensors placed on the side of the SHiNESS detector (bottom) for a $e^+e^-$ pair produced by a HNL decay. The red dot corresponds to the interaction vertex.}
\end{figure}

Figure \ref{fig:cherenkov_event} shows the spatial distribution for Cherenkov and scintillation photons and the relative charge collected by the photosensors in the SHiNESS detector for a $e^+e^-$ pair produced by a HNL decay. Isolating the Cherenkov component allows to reconstruct the direction of the two leptons, which would not be possible using only the scintillation light. 

\begin{figure}[htbp]
\begin{subfigure}[t]{0.49\textwidth}
\includegraphics[width=1\textwidth]{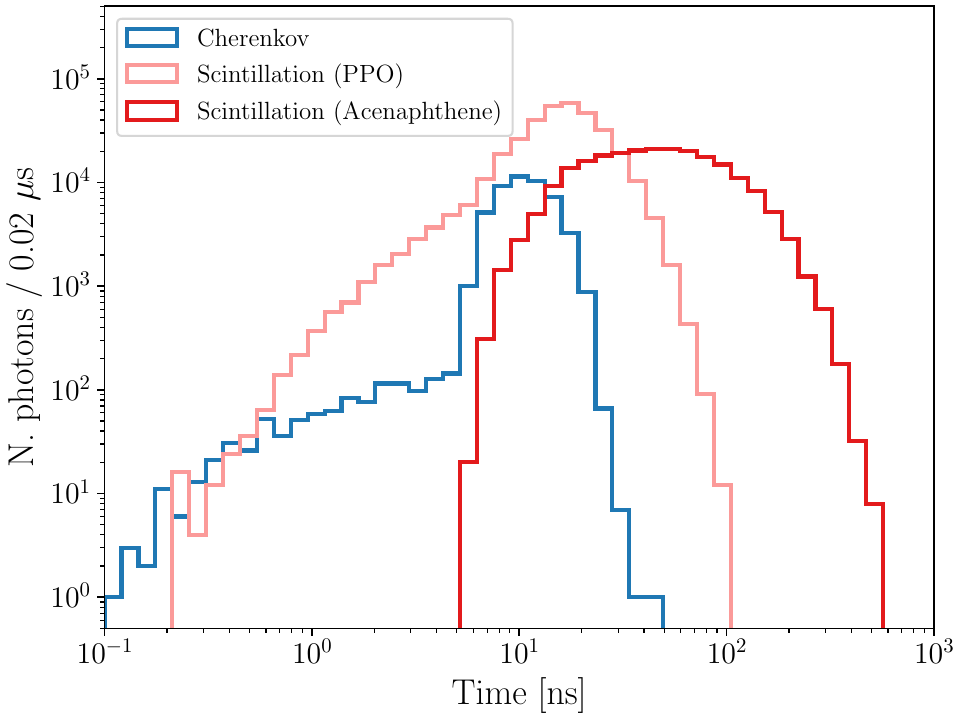}
\caption{\label{fig:cherenkov_time} Light emission time profile for Cherenkov and scintillation photons.}
\end{subfigure}
\hfill
\begin{subfigure}[t]{0.49\textwidth}
\includegraphics[width=1\textwidth]{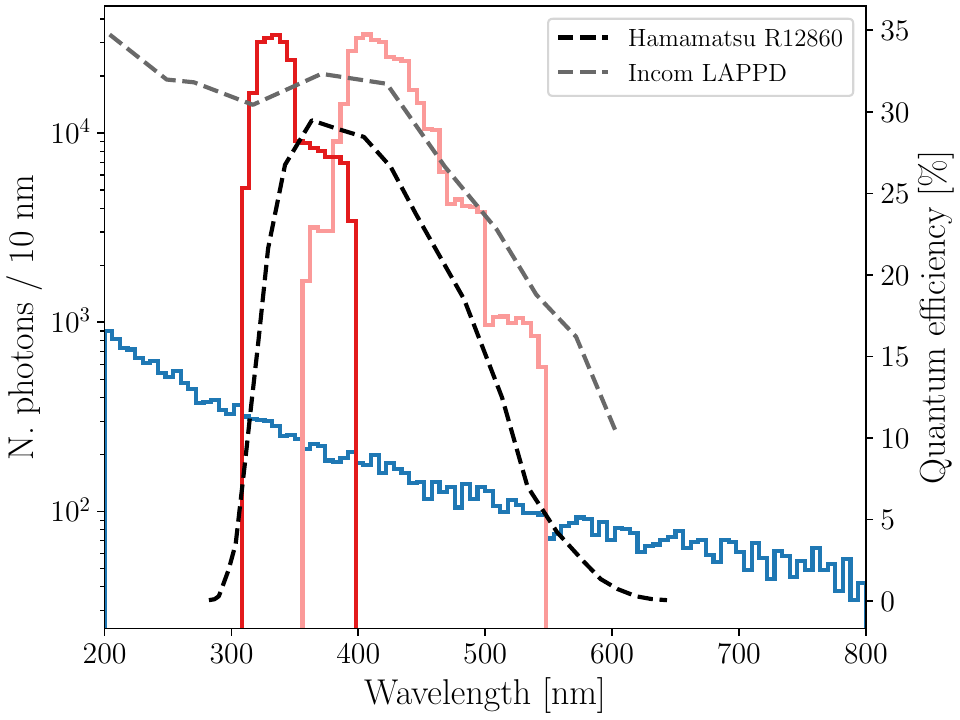}
\caption{\label{fig:cherenkov_wavelength} Wavelength spectrum for Cherenkov and scintillation photons (left axis) and QE for the two kinds of photosensor (right axis).}
\end{subfigure}
\caption{\label{fig:cherenkov} Scintillation with PPO (light red), scintillation with acenaphthene (red), and Cherenkov (blue) photon distributions as a function of time (left) and wavelength (right) for a 30~MeV electron in the SHiNESS detector. Quantum efficiency (QE) for the Hamamatsu PMT and the Incom LAPPD are also provided.}
\end{figure}

The time discrimination between Cherenkov and scintillation light could eventually be increased by loading the scintillator with \emph{slow fluors}, as detailed in ref.~\cite{Biller:2020uoi}. Figure \ref{fig:cherenkov_time} shows the arrival time distributions for the photons generated by a 30~MeV electron in the SHiNESS detector, using PPO or acenaphthene, one of the slow fluor candidates. In the latter case, the separation in time between Cherenkov and scintillation photons increases up to few ns. After requiring a Cherenkov light detection efficiency of 50\%, the purity of the Cherenkov selection goes from 23\% in the case of PPO to 91\% in the case of acenaphthene.

The R12860 and LAPPD quantum efficiencies are well matched to the scintillation and Cherenkov emission spectrum (see figure \ref{fig:cherenkov_wavelength}). However, absorption length of LAB/PPO increases rapidly in the UV, so only a small portion of the UV Cherenkov photons will reach the photosensors. 
A rich R\&D program is devoted at increasing the discrimination between scintillation and Cherenkov photons in liquid scintillator~\cite{Wonsak:2018uby, Aberle:2013jba}. A notable example is represented by quantum dots (QDs): these are semiconducting nanocrystals that can be loaded in the liquid scintillator and can shift the UV Cherenkov photons towards the visible with a high efficiency \cite{Gruszko:2018gzr}. In the SHiNESS detector, a simplified simulation of perovskite quantum dots loading \cite{plasmachem} gives an improvement of approximately a factor of 1.7 in the number of detected Cherenkov photons. Table~\ref{tab:rd} shows a summary of the possible cocktails for the liquid scintillator and the relative Cherenkov detection performances. 

\begin{table}[htbp]
\centering
\begin{tabular}{l|ll}
\hline
Cocktail &
\specialcell{Cherenkov sample purity\\at 50\% efficiency} &
\specialcell{Detected Cherenkov photons \\with $\Delta t<5$~ns (median)} \\
\hline
LAB/PPO & 23\% & 10.4 \\
LAB/Acenaphthene & 91\% & 10.4 \\
LAB/PPO + QDs & 23\% & 18.2 \\
\hline
\end{tabular}
\caption{\label{tab:rd}%
Purity of the sample of detected Cherenkov photons, requiring a 50\% efficiency, and median number of detected Cherenkov photons in the first 5 ns for various liquid scintillator cocktails. The numbers were obtain by simulating a sample of positrons produced by IBD events in the SHiNESS detector.}
\end{table}

Alternatively, several collaborations are developing water-based liquid scintillators (WbLS)~\cite{ANNIE:2017nng, Theia:2019non, WATCHMAN:2015lcq}, which uniquely combine the high light yield and low threshold of scintillator with the directionality of a Cherenkov detector~\cite{Alonso:2014fwf}. 
As an example, the THEIA collaboration~\cite{Theia:2019non} is exploring the separation of Cherenkov and scintillation photons for direction reconstruction of few-MeV electrons using a WbLS cocktail.

\section{\label{sec:pheno} Phenomenological implications of an extended neutrino sector}

As outlined in the introduction, an extended neutrino sector is well-motivated as the SM particle content needs to be enlarged in order to generate neutrino masses. Depending on the nature of the new particles and the values of their masses, different phenomenological consequences arise. We start discussing the implications of a non-unitary leptonic mixing matrix in section~\ref{sec:unitarity}, as it provides a model-independent way to search for indirect evidence of new states in the neutrino sector, that is valid across a wide range of masses and for a plethora of models. 

However, SHiNESS will also be able to search for direct signals due to additional neutrinos, either through the observation of oscillation signals (if there are new neutrinos present at the eV scale) or from their decays (for heavy neutrinos, with masses around the MeV scale or above). These two scenarios are discussed separately, in sections~\ref{sec:lightsterile} and~\ref{sec:hnl}, respectively.

\subsection{\label{sec:unitarity} Probing the unitarity of the lepton matrix}

Neutrino oscillations generally assume a unitary $3 \times 3$ lepton mixing matrix $U$. In the Pontecorvo–Maki–Nakagawa–Sakata parametrization this matrix is written by means of three mixing angles and one complex (Dirac) CP phase~\cite{ParticleDataGroup:2022pth}, plus two additional CP phases if neutrinos are Majorana. The unitarity of the matrix, however, only holds in a limited amount of neutrino mass models. In general, in the presence of $n$ additional neutrinos, the mass Lagrangian in the extended neutrino sector is diagonalized by a $(n+3)\times(n+3)$ mixing matrix:
\begin{equation}
\mathcal{U} = 
        \begin{pmatrix*}[l]
            U & W \\
            Z & V \\
        \end{pmatrix*},
\end{equation}
where $V$ is a $n\times n$ matrix and $W$ and $Z$ are $3 \times n$ and $n \times 3$ rectangular matrices, respectively. Here, $U$ is the $3\times 3$ active neutrino block of the matrix, which does not have to be unitary anymore. In the simple case of a type-I Seesaw~\cite{Minkowski:1977sc,Mohapatra:1979ia,Yanagida:1979as,Gell-Mann:1979vob}, the presence of right-handed heavy neutrinos causes unitarity deviations for $U$ which are proportional to the ratio between the masses of the light and heavy states. While a naive type-I seesaw would yield an exceedingly small effect ($\mathcal{O}(10^{-13})$ for $\mathcal{O}(\mathrm{TeV})$ heavy states), there are several examples of neutrino mass models predicting larger deviations~\cite{Mohapatra:1979ia, Antusch:2014woa, Nelson:2007yq, Antusch:2006vwa}. Thus, searching for unitarity violation represents a model-independent way to probe for non-standard neutrino states and interactions~\cite{Blennow:2016jkn}, with direct sensitivity to BSM physics. Moreover, if the mixing matrix in the active block, $U$, is not unitary, then additional CP phases would enter its definition. In fact, deviations from unitarity in the active block are often described using a lower triangular parametrization~\cite{Xing:2007zj, Xing:2011ur, Escrihuela:2015wra}, which includes three off-diagonal complex parameters, each of them with an associated CP phase. Therefore, non-unitarity effects could a priori be relevant for the upcoming neutrino oscillation experiments DUNE and Hyper-Kamiokande, which will aim to determine whether there is CP violation in the lepton sector.

A $3\times 3$ unitary matrix should meet a total of 9 conditions, which are easily derived from the requirement that $U^{\dag}U=UU^{\dag}=\mathbb{I}$. They boil down to requiring that all its rows are normalized to one, and that the so-called \emph{unitarity triangles} close:
\begin{eqnarray}
\label{eq:unitarity_row}
r_{\alpha} & \equiv &  |U_{\alpha 1}|^2 + |U_{\alpha 2}|^2 + |U_{\alpha 3}|^2 = 1  \, , \\
t_{\alpha \beta} & \equiv & U_{\alpha 1}^*U_{\beta 1} + U_{\alpha 2}^*U_{\beta 2} + U_{\alpha 3}^*U_{\beta 3} = 0 \quad (\alpha \neq \beta)\, ,
\label{eq:unitarity_triang}
\end{eqnarray}
where $\alpha,\beta$ indicate active neutrino flavor indices $(e, \mu, \tau)$. If $U$ is not unitary then the conditions in eqs.~\eqref{eq:unitarity_row}, \eqref{eq:unitarity_triang} do not have to be satisfied. Thus, a way to constrain the unitarity of the leptonic mixing matrix consists in probing the quantities $r_\alpha$ and $t_{\alpha\beta}$ defined above. In this case it is worth noting though, that if the whole matrix $\mathcal{U}$ is unitary, then the application of Cauchy-Schwarz inequalities imposes restrictions on the values of $r_\alpha$ and $t_{\alpha\beta}$. This allows to place indirect limits on unitarity deviations which can in some cases be stronger than direct bounds~\cite{Antusch:2006vwa,Parke:2015goa}. However, in order to be as general as possible, here we adopt instead an \emph{agnostic} assumption (as in ref.~\cite{Ellis:2020hus}) and do not assume that $\mathcal{U}$ is unitary. 

Experimentally, a non-unitary matrix leads to a non-vanishing probability to have flavor transitions at negligible distances, also known as \emph{zero distance} effects~\cite{Antusch:2006vwa}. The oscillation probabilities in this case read
\begin{align}
    P(\nu_{\alpha}\rightarrow\nu_{\alpha}) & = r_\alpha^2 \, ,\\
    P(\nu_{\alpha}\rightarrow\nu_{\beta}) & = |t_{\alpha \beta}|^2 \, .
\end{align}
A short-baseline neutrino experiment such as SHiNESS may therefore be sensitive to such effects since standard, three-flavor oscillations have yet to develop. From the expressions above, however, it is evident that the sensitivity to $r_\alpha$ will be directly limited by the systematic uncertainties associated to the normalization of the event sample and therefore will not be discussed here. A search for anomalous appearance is much more promising, provided that backgrounds can be sufficiently reduced. In particular, SHiNESS is expected to be sensitive to the closure of the electron-muon triangle parameter $t_{e \mu}$ through the appearance probability $P(\bar{\nu}_e\rightarrow\bar{\nu}_{\mu})$, looking for an excess of IBD events in the detector.

\subsection{\label{sec:lightsterile} Light sterile neutrinos}

If the extra neutrinos are light enough, they could participate in oscillations, leading to a new oscillation frequency and a smoking signal at neutrino detectors. As outlined in the introduction, throughout the past two decades a plethora of experiments have reported contradicting signals in the context of sterile neutrino searches in oscillation experiments~\cite{Diaz:2019fwt, Dentler:2018sju}. If the reported anomalies are caused by the existence of a fourth mass neutrino eigenstate, the best value for the mass splitting $\Delta m_{41}^2 = m_4^2-m_1^2$ is around the $\mathcal{O}(\mathrm{eV}^2)$ scale~\cite{Dentler:2018sju} and much larger than the other two mass splittings $\Delta m_{31}^2$, $\Delta m_{21}^2$. For this reason this is usually referred to as a \emph{(3+1) scenario}. For sufficiently short baselines, $\Delta m_{21}^2L/E_{\nu} \ll 1$ and $\Delta m_{31}^2L/E_{\nu} \ll 1$, so that standard three-flavor neutrino oscillations have yet to develop. In this limit, disappearance and appearance oscillations can be approximated as:
\begin{align}
    P(\nu_\alpha \to \nu_\alpha) &\simeq 1 - \sin^2\left(2\theta_{\alpha\alpha}\right) \sin^2\left(\frac{\Delta m_{41}^2 L}{4E_\nu}\right) \, , \label{eq:Pee} \\
    P(\nu_\alpha \to \nu_\beta) &\simeq \sin^2\left(2\theta_{\alpha \beta}\right) \sin^2\left(\frac{\Delta m_{41}^2 L}{4E_\nu}\right) \, , \quad \label{eq:Pme} 
\end{align}
where $E_{\nu}$ is the neutrino energy, $L$ is the distance traveled by the neutrino~\cite{Dasgupta:2021ies}, and the last expression only holds for $\alpha \neq \beta$. Here we have adopted the usual definitions for the effective mixing angles $\theta_{\alpha\alpha}, \theta_{\alpha\beta}$ that control the amplitude of the oscillation (which eventually can be mapped onto the elements of the full mixing matrix, $\mathcal{U}_{\alpha 4}$ and $\mathcal{U}_{\beta 4}$).

The SHiNESS experiment has the ability of definitely clarifying the complex (and, at times, contradictory) experimental scenario revolving the existence of a sterile neutrino, thanks to its key features:
\begin{itemize}
    \item will detect $\pi^+$ DAR neutrinos, whose spectrum is well understood and shape systematic uncertainties are small;
    \item will sit at a distance from the neutrino source similar to the one of the LSND experiment (25~m vs. 30~m). This will allow to cover the entire parameter space in the (3+1) model allowed by the LSND result, complementary to the space covered by the JSNS$^2$ experiment~\cite{JSNS2:2013jdh};
    \item is sensitive to both the $\bar{\nu}_{\mu}\rightarrow\bar{\nu}_e$ appearance channel and the $\nu_e\rightarrow\nu_e$ disappearance channel, thus probing the light sterile neutrino model for both the LSND/MiniBooNE and the gallium anomaly.
\end{itemize}

We finalize this discussion by stressing the similarities between the phenomenology associated to a non-unitary mixing matrix and that of an averaged-out sterile neutrino oscillation. Generally speaking, oscillations are said to be averaged-out when the corresponding oscillation frequency leads to oscillations that are too fast to be resolved by the detector, considering its energy and spatial resolution as well as the distance to the detector and the energy of the neutrinos. In the case of SHiNESS this translates into the condition $\Delta m_{41}^2 \gtrsim 20~\mathrm{eV}^2$. As discussed in refs.~\cite{Blennow:2016jkn, Coloma:2021uhq}, at leading order the phenomenological consequences of a non-unitarity mixing matrix are mostly equivalent to those of an averaged-out sterile neutrino, up to quartic terms in the mixing between the heavy and active states (see refs.~\cite{Fong:2016yyh,Fong:2017gke} for a discussion on such effects). Thus, above this value the limits on the unitarity of the lepton mixing matrix (obtained in a model-independent way) can be mapped onto the limits on the mixing angle for a sterile neutrino (and \emph{viceversa}). Conversely, for sterile neutrino masses below this scale the corresponding oscillations could be resolved and measured at the detector. As the two limits yield very different phenomenological implications, we will study them separately.

\subsection{\label{sec:hnl} Heavy Neutral Leptons}

If the new neutrino states have their masses at the MeV or GeV scale, they are usually refereed to as Heavy Neutral Leptons (HNL). HNLs can couple to SM particles through their mixing with the SM neutrinos and therefore may be produced copiously in meson decays as well as muon or tau decays. Once produced, a HNL may travel for hundreds or thousands of meters before decaying back to a combination of SM leptons and mesons. This makes fixed-target experiments an ideal setup to search for their decay signals (for a recent review of the strongest constraints on HNLs see e.g. ref.~\cite{Fernandez-Martinez:2023phj}). Specifically, the large number of stopped pions and muons at spallation sources makes them outstanding facilities to search for HNLs below the pion mass~\cite{Ema:2023buz}. 

Once produced near the ESS beam target, HNLs can be long-lived enough to reach the SHiNESS detector and decay visibly inside as $N \to \nu_\alpha e^+e^-$, where the flavor of the neutrino in the final state will depend on the mixing of the HNL with the active neutrinos. Hereafter we follow the analysis of ref. \cite{Ema:2023buz} and consider two cases separately, depending on whether the HNL mixes predominantly with muon or electron neutrinos. 

In the muon mixing case, the HNL is mainly produced from the decay of muons, although for sufficiently light masses ($m_N < 
m_{\pi}-m_{\mu}$) it may also be produced subdominantly from pion decays. The HNL then decays via the neutral current, with the following decay width:
\begin{align}
	\Gamma(N \to e^+ e^- \nu_\mu) = \frac{G_F^2 m_N^5}{768\pi^3}\abs{U_{\mu N}}^2
	\left(1-4\sin^2\theta_W + 8\sin^4\theta_W\right),
\end{align}
where $\theta_W$ is the Weinberg angle.

In the electron mixing case, the HNL is abundantly produced both from the decay of muons and  pions. Contrary to the muon mixing case, in this case the HNL can decay both via the neutral current and the charged current. The corresponding decay rate reads~\cite{Gorbunov:2007ak, Atre:2009rg}:
\begin{align}
	\Gamma(N \to e^+ e^- \nu_e)
	&= \frac{G_F^2 m_N^5}{768\pi^3}\abs{U_{eN}}^2\left(1 + 4\sin^2\theta_W + 8\sin^4\theta_W\right).
\end{align}

In the limit of a very long-lived HNL of mass $m_{N}$, the total number of HNL decays in the detector can be obtained approximately as:
\begin{align}
	N^{|U_{lN}^2|}_{ee,i} & \simeq N_i 
 	\times
    \epsilon_\mathrm{det}^{(i)}(m_N)
    \times
    \epsilon_\mathrm{acc}
	\times \frac{1}{\Gamma_i}\int dE_N
	\frac{d\Gamma(i \to l N )}{dE_N}
    \times
	\frac{L_\mathrm{det}}{\gamma\beta c\tau_{N}} \times \mathrm{BR}(N\to \nu ee),
	\label{eq:Nee_total}
\end{align}
where $N_i$ stands for the total number of parent particles ($i=\mu, \pi$), $f$ represents the other particles produced together with the HNL (which depend on the decay channel), $d\Gamma(i \to f N )/dE_N$ is the differential production rate as a function of the HNL energy $E_N$ (which can be found e.g. in ref.~\cite{Ema:2023buz}), and $\Gamma_i$ is the total decay width of particle $i$. The detector acceptance is defined by $\epsilon_\mathrm{acc}$ and the selection efficiency is given by $\epsilon_\mathrm{det}^{(i)}(m_N)$, see figure~\ref{fig:hnl_eff}. When the parent particle $i$ is a muon the HNL production is not monochromatic, thus $\epsilon_\mathrm{det}^{(i)}(m_N)$ is taken as the weighted average according to the HNL energy spectrum.  The length of the detector is $L_{\mathrm{det}}$, and $c\beta\tau_{N}$ is the decay length of the HNL in the lab frame, while $\mathrm{BR}(N \to \nu e e )$ is the branching ratio into the decay channel $N\to\nu ee$. It follows from eq.~\eqref{eq:Nee_total} that, in this limit, the result only depends on $\Gamma(N \to \nu e e)$, while the dependence on the total width of the HNL effectively drops out.

\section{\label{sec:principle}Expected number of signal events}

\subsection{Neutrino detection}
\label{sec:osc-sig}
\subsubsection{Inverse beta decay}\label{sec:ibd}

\begin{figure}[htbp]
\begin{subfigure}[t]{0.48\textwidth}
    
\includegraphics[width=1\textwidth]{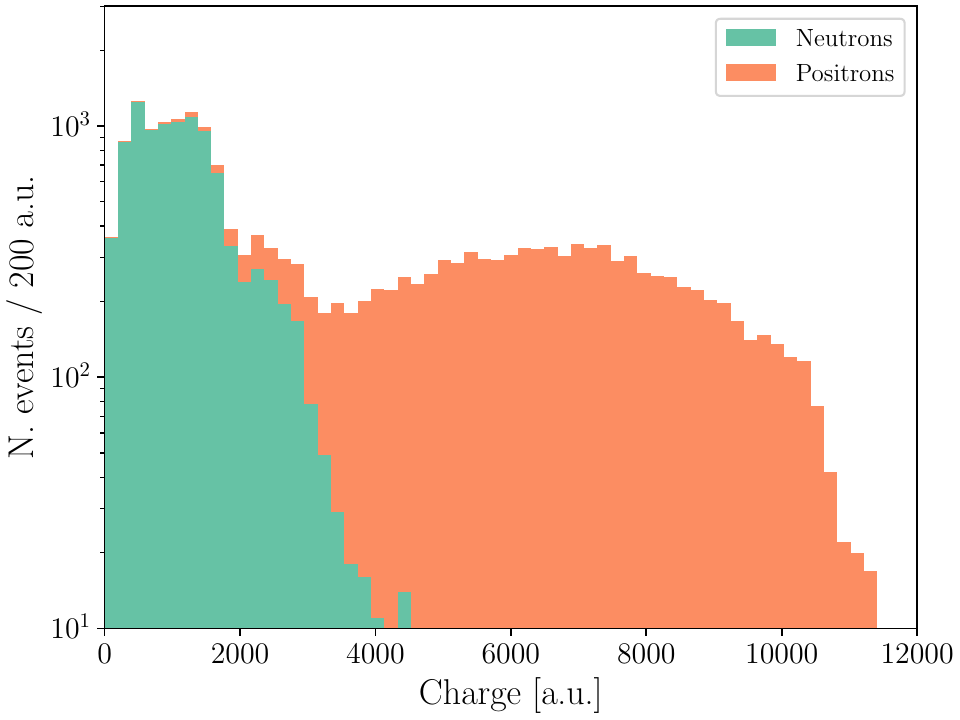}
\caption{Stacked histogram of the sum of the charge collected by the PMTs for each event.}
\end{subfigure}
\hfill
\begin{subfigure}[t]{0.48\textwidth}

\includegraphics[width=1\textwidth]{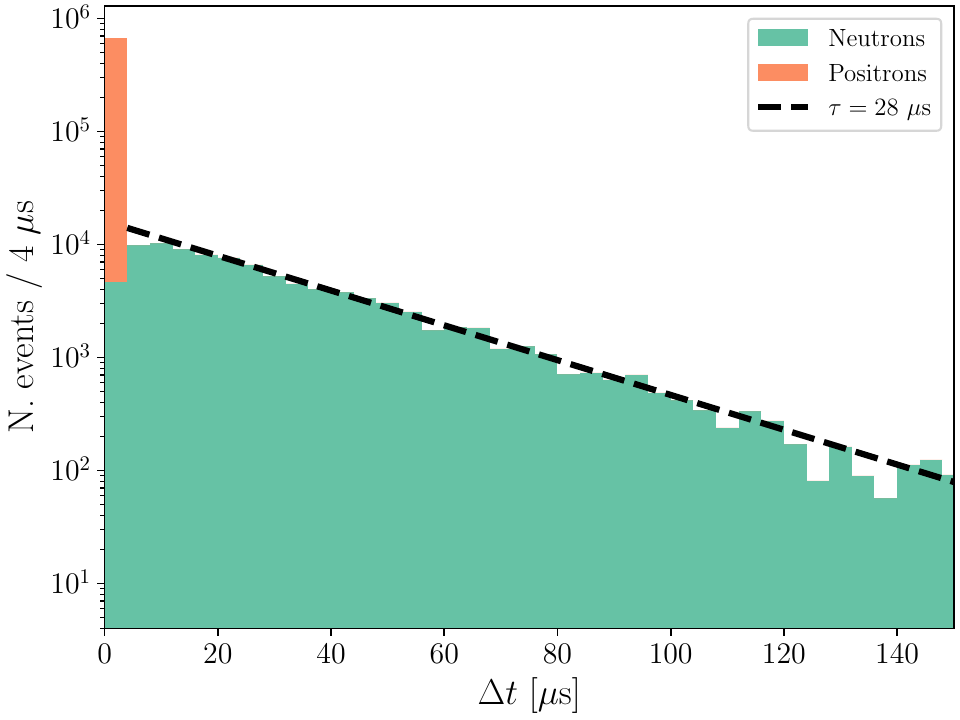}
\caption{Stacked histogram of the time difference between the IBD interaction and the signal generated by each particle.}
\end{subfigure}
\caption{\label{fig:electron_spectrum}Charge and time distribution for $10^4$ IBD events in the SHiNESS detector. The positrons produce a prompt signal with a broad charge distribution, while the neutrons are absorbed with a characteristic time constant of around 28~\unit{\micro\second} and are concentrated at lower energies, corresponding to the 8~MeV capture by Gd and the 2.2~MeV capture by hydrogen.}
\end{figure}

The interaction of $\bar{\nu}_{e}$ in the liquid scintillator can be detected through to observation of the IBD process both on protons:
\begin{equation}
    \bar{\nu}_{e} + p \rightarrow e^+ + n.
\end{equation}
and on carbon atoms:
\begin{equation}
    \bar{\nu}_{e} + ^{12}\mathrm{C} \rightarrow e^+ + n + ^{11}\mathrm{B}.
\end{equation}
The experimental signature of this process is the emission of a positron followed by the delayed neutron capture and its $\gamma$ emission. Figure \ref{fig:electron_spectrum} shows the time and charge distribution for $10^4$ IBD events in the SHiNESS detector, obtained with a Geant4 simulation. The positron generates a \emph{prompt} signal, while the neutron absorption generates a \emph{delayed} signal with a characteristic exponential time constant of around 28~\unit{\micro\second}. Neutrons generate signal in the lower end of the charge spectrum, while the positrons follow a broad Michel distribution. 

Assuming an active mass of 42~ton of liquid scintillator and knowing the flux-averaged cross sections we can calculate the number of inverse beta decay events we expect in case of $\bar{\nu}_e$ appearance. 
The yearly flux of $\bar{\nu}_{\mu}$ from $\pi^+$ DAR that can oscillate into $\bar{\nu}_e$ at a distance of $L=$25~m from the beam target is given by:
\begin{align}\label{eq:flux}
    \phi_{\bar{\nu}_{\mu}} = &\frac{8.5\times10^{22}}{4\pi L^2}~\mathrm{events/year} \\= &1.08 \times 10^{15}~\mathrm{cm}^{-2}~\mathrm{events/year}. 
\end{align}
The total number of oscillated $\bar{\nu}_e$ events, assuming a 100\% oscillation probability and a 10\% systematic uncertainty, is given by the sum of the interactions on the proton and carbon targets:
\begin{align}\label{eq:ibd}
    N_{\bar{\nu}_{e}}^{\mathrm{IBD}} = ( &N_\mathrm{H}\cdot\sigma_{\bar{\nu}_{e}p \rightarrow e^+n}^{\bar{\nu}_{\mu}} + \\
    & N_{\mathrm{C}}\cdot\sigma_{\bar{\nu}_{e}^{12}\mathrm{C} \rightarrow e^+n^{11}\mathrm{B}}^{\bar{\nu}_{\mu}} ) \cdot\phi_{\bar{\nu}_{\mu}} \cdot \epsilon_{\mathrm{det}} \nonumber \\
    =& (1.48\pm0.15) \times 10^5 ~\mathrm{events/year}, \nonumber
\end{align}
where:
\begin{itemize}
\item $N_\mathrm{H} = 2.76\times10^{30}$ is the number of hydrogen targets;
\item $N_\mathrm{C}=1.70\times10^{30}$ is the number of carbon targets;
\item $\sigma_{\bar{\nu}_{e}p \rightarrow e^+n}^{\bar{\nu}_{\mu}}=93.5 \times 10^{-42}$~cm$^2$ is the $\bar{\nu}_{\mu}$ flux-averaged cross section for the IBD process on proton~\cite{KARMEN:2002zcm, Vogel:1999zy};
\item $\sigma_{\bar{\nu}_{e}^{12}\mathrm{C} \rightarrow e^+n^{11}\mathrm{B}}^{\bar{\nu}_{\mu}}=8.5 \times 10^{-42}$~cm$^2$ is the $\bar{\nu}_{\mu}$ flux-averaged cross section for the IBD process on carbon~\cite{KARMEN:2002zcm, Kolbe:2000np};
\item $\epsilon_{\mathrm{det}}=0.5$ is the assumed SHiNESS detector efficiency. 
\end{itemize}

\subsubsection{Charged-current interaction}\label{sec:cc}
\begin{figure}[htbp]
\centering
\includegraphics[width=0.7\textwidth]{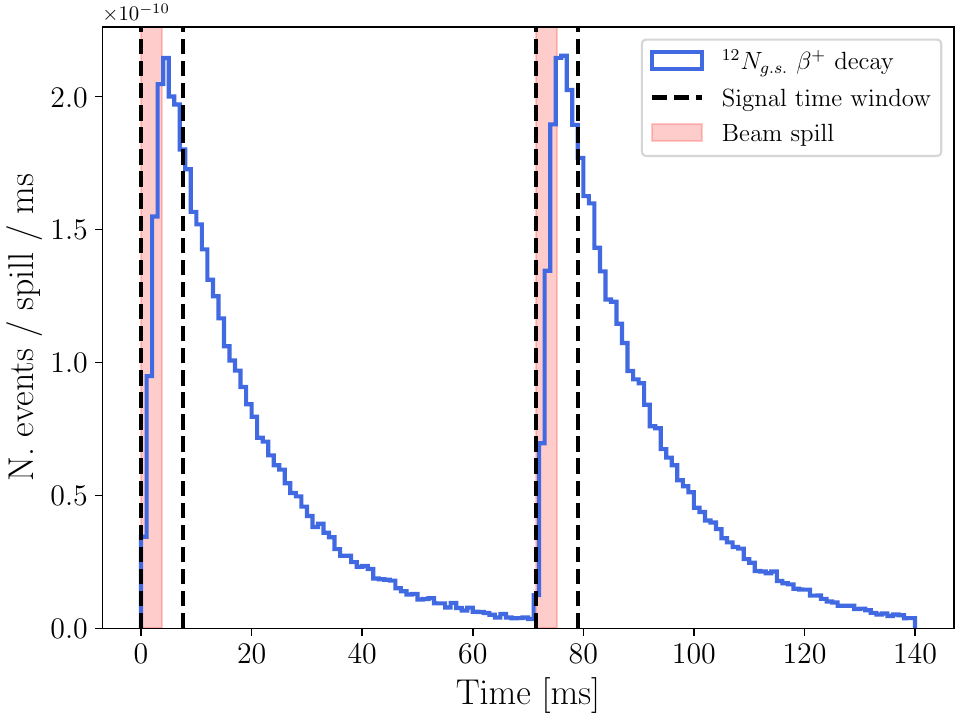}
\caption{\label{fig:nuebeam} Time distribution of $\nu_{e} + ^{12}\mathrm{C} \rightarrow e^{-}  + ^{12}\mathrm{N_{g.s.}}$ events (solid blue line) for two consecutive beam spills. The beam-spill time window is shown in light red and the $\nu_e$ disappearance signal time window is defined by the dashed black lines. The fraction of events inside the signal time window is 30\%.}
\end{figure}

The $\nu_e$ can interact on carbon via the charged-current channel:
\begin{align}\label{eq:cc}
    \nu_e + ^{12}\mathrm{C} \rightarrow &^{12}\mathrm{N_{gs}} + e^- \\
                               &^{12}\mathrm{N_{gs}} \rightarrow ^{12}\mathrm{C} + e^+ + \nu_e. \nonumber
\end{align}
The experimental signature of this process is the emission of the electron and the subsequent $^{12}\mathrm{N_{gs}}$ beta decay in delayed coincidence~\cite{Conrad:2011ce}. The scattered electron has a maximum kinetic energy $E_{e^-}$ of 35.5 MeV due to the $Q$ value of 17.3~MeV. The recoil energy of the $^{12}\mathrm{N_{gs}}$ nucleus is negligible, so $E_{\nu_e} = E_{e^-} + 17.3$~MeV~\cite{LSND:2001fbw}. The positron has a maximum kinetic energy of 16.3~MeV and the decay has a lifetime of 15.9 ms~\cite{Krakauer:1991rf}, which is significantly longer than the beam spill. 
Thus, in order to reduce the amount of steady-state backgrounds, we apply a timing cut of 2 beam spills (7.6~ms), which has an efficiency $\epsilon_{\mathrm{time}}=30\%$. Figure \ref{fig:nuebeam} shows the time distribution of the $^{12}\mathrm{N_{gs}}$ $\beta^+$-decay for two consecutive beam spills.

The yearly $\phi_{\nu_e}$ flux from $\pi^+$ DAR is the same as the $\phi_{\bar{\nu}_{\mu}}$ one of eq.~\eqref{eq:flux}, given the production process of eq.~\eqref{eq:dar}. The number of $\nu_e$ yearly charged-current interactions in the detector with an energy larger than 17.3~MeV, assuming a 10\% systematic uncertainty, is then given by:
\begin{align}
        N_{\nu_{e}}^{\mathrm{CC}} =& N_{\mathrm{C}}\cdot\sigma_{\nu_{e}^{12}\mathrm{C} \rightarrow e^{-} {^{12}}\mathrm{N_{gs}}}\cdot\phi_{\nu_e}(E_{\nu_e}>17.3~\mathrm{MeV})\cdot\epsilon_{\mathrm{det}}\cdot\epsilon_{\mathrm{time}} \\
        = &(2.19\pm0.22) \times 10^3 ~\mathrm{events/year}, \nonumber
\end{align}
where $\sigma_{\nu_{e}^{12}\mathrm{C} \rightarrow e^{-} {^{12}}\mathrm{N_{gs}}}=8.9\times 10^{-42}$~cm$^2$ is the flux-averaged cross section as calculated in ref.~\cite{Kolbe:1999au}. Both the KARMEN~\cite{KARMEN:1994xse} and the LSND experiment~\cite{LSND:2001fbw} found this calculation in good agreement with the data.

\subsubsection{Neutral-current interaction}\label{sec:nc}
All the three neutrino types $\bar{\nu}_{\mu}$, $\nu_{\mu}$ and $\nu_e$ can interact with the carbon atom through the neutral current (NC) channel and the subsequent emission of 15.11 MeV $\gamma$:
\begin{align}\label{eq:nc}
     ^{12}\mathrm{C} + \nu \rightarrow &^{12}\mathrm{C}^* + \nu \\
                              &^{12}\mathrm{C}^*\rightarrow^{12}\mathrm{C} + \gamma.\nonumber
\end{align}

The flux-averaged cross sections for $\nu_e$ and $\bar{\nu}_{\mu}$ and for the monoenergetic $\nu_{\mu}$ were calculated in ref.~\cite{Kolbe:1994xb} as $\sigma^{\nu_e, \bar{\nu}_{\mu}}_{^{12}\mathrm{C}\nu\rightarrow^{12}\mathrm{C}^*\nu} = 10.5 \times 10^{-42}$~cm$^2$ and $\sigma^{\nu_{\mu}}_{^{12}\mathrm{C}\nu\rightarrow^{12}\mathrm{C}^*\nu} = 2.8 \times 10^{-42}$~cm$^2$, respectively. Both cross sections were measured by the KARMEN experiment 
and found in good agreement with the calculation~\cite{KARMEN:1995dwo, KARMEN:1998xmo}. 
Given the large beam spill of the ESS, it is not possible to distinguish between $\nu_{\mu}$, produced by prompt $\pi^+$ decay ($\tau = 26$~ns), and $\nu_e$ and $\bar{\nu}_{\mu}$, produced by slower $\mu^+$ decay ($\tau = 2.2$~\si{\micro\second}).
Thus, we consider a total cross section for the entire flux $\sigma_{^{12}\mathrm{C}\nu\rightarrow^{12}\mathrm{C}^*\nu}=13.3\times 10^{-42}$~cm$^2$.

The total number of neutral current $\pi^+$ DAR neutrino interactions on carbon, assuming a 10\% systematic uncertainty, is then given by:
\begin{align}
    N_{\nu}^{\mathrm{NC}} =& N_{\mathrm{C}}\cdot\sigma_{^{12}\mathrm{C}\nu\rightarrow^{12}\mathrm{C}^*\nu}\cdot\phi_{\nu}\cdot\epsilon_{\mathrm{det}} \\
        = &(7.33\pm0.73) \times 10^4 ~\mathrm{events/year}, \nonumber,
\end{align}
where $\phi_{\nu} = 3.24\times10^{15}$~cm$^{-2}$ events/year is the sum of the $\nu_e$, $\nu_{\mu}$, and $\bar{\nu}_{\mu}$ fluxes.

\subsection{Decays of Heavy Neutral Leptons\label{sec:hnl-sig}}

\begin{figure}[htbp]
\centering
\includegraphics[width=0.7\textwidth]{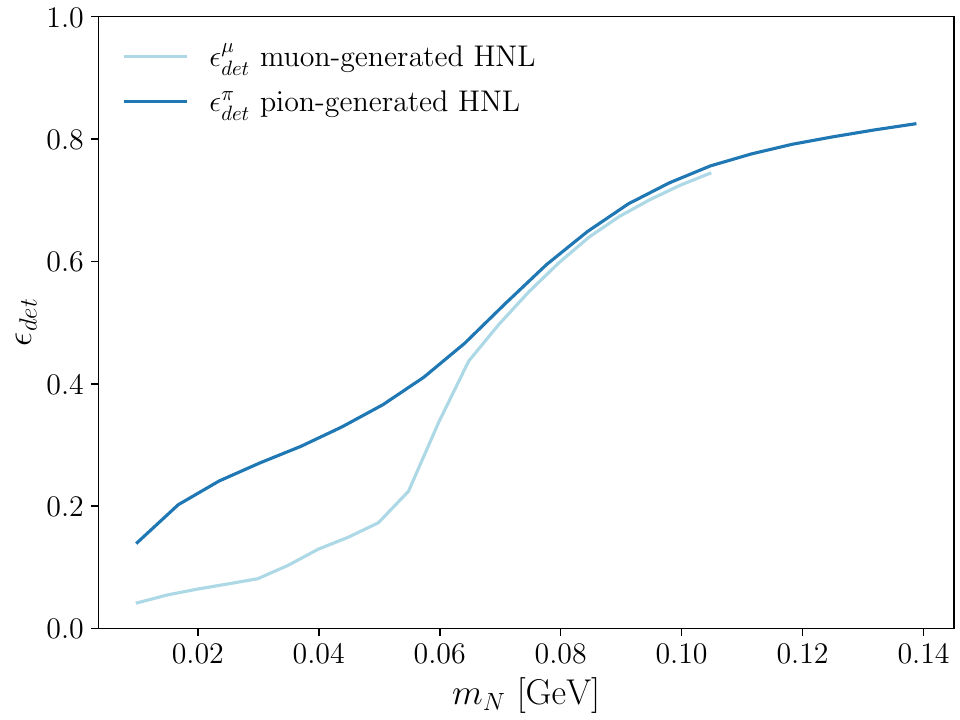}
\caption{Selection efficiency in the SHiNESS detector for HNLs in the case of pion (blue) and muon (light blue) production. The angle of the $e^+e^-$ pair is required to be larger than $15^{\circ}$ and the reconstructed energy of each particle is required to be larger than 17~MeV.}\label{fig:hnl_eff} 
\end{figure}

As outlined in section~\ref{sec:hnl}, the main signal expected at SHiNESS from a HNL with mass below the pion mass is an excess of events with an electron positron pair in the final state (plus a neutrino, which of course exists the detector unobserved). The expected number of signal events can be estimated inserting the values for SHiNESS into eq.~\eqref{eq:Nee_total}, for sufficiently low values of the mixing (so that the approximation of a very long-lived particle holds). For the electron-mixing case we obtain:
\begin{align}
	N_{ee,\mu}^{|U_{eN}|^2} & \simeq
	\displaystyle 3.8\times 10^{6}\left(\frac{\abs{U_{eN}}^2}{10^{-6}}\right)^2
	\left(\frac{m_N}{m_\mu}\right)^6 \left(1-\frac{m_N}{m_\mu}\right)^4
	\left(1 + \frac{4m_N}{m_\mu} + \frac{m_N^2}{m_\mu^2}\right) \, , \\
	N_{ee,\pi}^{|U_{eN}|^2} & \simeq 1.9\times 10^{6}\left(\frac{\abs{U_{eN}}^2}{10^{-6}}\right)^2
	\left(\frac{m_N}{m_\mu}\right)^7
	\frac{m_N \, m_\pi \, (m_\pi^2 - m_N^2)}{(m_\pi^2 - m_\mu^2)^2} \, ,
\end{align}
where in the last expression we have neglected the mass of the electron for simplicity. In the muon-mixing scenario, on the other hand, we get:
\begin{align}
	N_{ee,\mu}^{|U_{\mu N}|^2} & \simeq
	6.9\times 10^{5}\left(\frac{\abs{U_{\mu N}}^2}{10^{-6}}\right)^2
	\left(\frac{m_N}{m_\mu}\right)^6 \left(1-\frac{m_N}{m_\mu}\right)^4
	\left(1 + \frac{4m_N}{m_\mu} + \frac{m_N^2}{m_\mu^2}\right) \, ,\\
	N_{ee,\pi}^{|U_{\mu N}|^2} & \simeq
	4.0\times 10^{5} \left(\frac{\abs{U_{\mu N}}^2}{10^{-6}}\right)^2 \left(\frac{m_N}{m_\mu}\right)^6
	\frac{m_\pi \left[
	(m_\mu^2 + m_N^2) m_\pi^2 - (m_\mu^2 + m_N^2)^2 + 4m_\mu^2m_N^2
	\right]}{m_\mu(m_\pi^2-m_\mu^2)^2}  \, .
\end{align}
For each case, and for a given value of the mass and the mixing, the total number of events is obtained adding both contributions from pion and muon decays, multiplied by the corresponding selection efficiency:

Experimentally, the $e^+ e^-$ pair will be emitted in the laboratory frame with a certain angle, which is essentially controlled by $m_N$. The selection efficiency in the SHiNESS detector for HNL events produced by muons and pions decay is shown in figure \ref{fig:hnl_eff}, assuming an angular resolution of $10^{\circ}$. This value represents an educated guess on the capability of the detector and is comparable with the $\sim12^{\circ}$ of LSND \cite{Conrad:2013mka}, which only employed PMTs. A cut on the reconstructed energy of 17~MeV and a cut on the angle between the two leptons of $15^{\circ}$ are applied. The dependence on the parent particle comes from the different energy spectra, which leads to a slightly different boost and therefore a different opening angle for the electron pair. 

In the case of muon mixing, the pion can produce the HNL only if $m_{N} < m_{\pi} - m_{\mu}$. Thus, given the reconstructed energy cut, the efficiency in this case is effectively 0 and the number of events for the electron and muon mixing are given by:

\begin{align}
    N_{ee}^{|U_{eN}|^2} &= N_{ee,\mu}^{U_{eN}|^2} \times \epsilon_{\mathrm{det}}^{\mu} + N_{ee,\pi}^{|U_{eN}|^2} \times \epsilon_{\mathrm{det}}^{\pi},\\
    N_{ee}^{|U_{\mu N}|^2} &= N_{ee,\mu}^{|U_{\mu N}|^2} \times \epsilon_{\mathrm{det}}^{\mu}.
\end{align}

\section{\label{sec:bkg}Expected backgrounds}
\subsection{Backgrounds to neutrino detection\label{sec:osc-bg}}
\subsubsection*{\texorpdfstring{$\bar{\nu}_e$}{bar nu e} beam component}\label{sec:nue} 
Although the vast majority of $\mu^-$ is absorbed in the target before it can decay, a small fraction can decay in flight and emit $\bar{\nu}_e$, which are indistinguishable from the ones caused by the eventual $\bar{\nu}_{\mu}\rightarrow\bar{\nu}_e$ oscillation. A simulation of the ESS beam with the BERT-HP physics list gives a ratio $\phi_{\bar{\nu}_e}/\phi_{\bar{\nu}_{\mu}}=2.2\times10^{-3}$, after requiring the interaction to happen during the beam spill time window.The number of background events due to intrinsic $\bar{\nu}_e$ beam component assuming a 25\% systematic uncertainty is then given by:
\begin{align}\label{eq:nue_beam}
    N_{\bar{\nu}_{e}}^{\mathrm{beam}} = ( &N_\mathrm{H}\cdot\sigma_{\bar{\nu}_{e}p \rightarrow e^+n}^{\bar{\nu}_e} + \\
    & N_{\mathrm{C}}\cdot\sigma_{\bar{\nu}_{e}^{12}C \rightarrow e^+n^{11}\mathrm{B}}^{\bar{\nu}_e} ) \cdot\phi_{\bar{\nu}_e} \cdot \epsilon_{\mathrm{det}} \nonumber \\
    =& (254\pm65)~\mathrm{events/year}, \nonumber
\end{align}
where:
\begin{itemize}
     \item $\sigma_{\bar{\nu}_{e}p \rightarrow e^+n}^{\bar{\nu}_e}=72\times10^{-42}$~cm$^2$ is the $\bar{\nu}_e$ flux-averaged cross section for the IBD process on proton~\cite{KARMEN:2002zcm};
     \item $\sigma_{\bar{\nu}_{e}^{12}\mathrm{C} \rightarrow e^+n^{11}\mathrm{B}}^{\bar{\nu}_e}=7.4\times10^{-42}$~cm$^2$ is the $\bar{\nu}_e$ flux-averaged cross section for the IBD process on carbon~\cite{KARMEN:2002zcm}.
\end{itemize} 
This is an irreducible background which needs to be precisely assessed with a detailed simulation of the ESS beam.

\subsubsection*{Cosmic-ray backgrounds}\label{sec:cosmic}
Although steady-state backgrounds can be precisely quantified during beam-off periods, it is important to suppress them as much as possible to avoid losing signal sensitivity. Cosmic rays can be efficiently vetoed by scintillator panels surrounding the detectors, as in the KARMEN experiment~\cite{Eitel:2000by}, or by an outer layer of liquid scintillator equipped with PMTs, which is the solution chosen by the JSNS$^2$ experiment~\cite{JSNS2:2021hyk}. The main background caused by cosmic rays is represented by cosmic neutrons entering the detector, which can't be vetoed, and by fast neutrons produced by cosmic muons impinging on the materials surrounding the detector. These neutrons can enter the detector, recoil the protons being thermalized, and finally be captured, emitting 8 MeV gammas. They can mimic both the prompt signal (with the recoil proton) and the delayed coincidence (with the neutron absorption). They represent the dominant background of the JSNS$^2$ experiment~\cite{Maruyama:2022juu}.

A CORSIKA simulation~\cite{Heck:1998vt} of the cosmic-ray rate at the ESS site gives a cosmic neutron rate in the detector of 5.8~Hz, assuming a 30~cm concrete roof. This rate can be reduced by a factor of $10^3$ after applying conservative selection cuts on the energy, time, and position of the event in the detector, which translates into $1.6\times10^{-5}$ events/spill. 

A further reduction in the number of selected events can be obtained by applying particle identification (PID) techniques. The LSND collaboration pioneered a method which combines scintillation and Cherenkov light emissions and achieves a rejection factor of around 100~\cite{LSND:2001aii}. 

The JSNS$^2$ collaboration has developed a pulse-shape discrimination algorithm by dissolving 10\% Di-Isopropyl-Naphthalene (DIN) by weight into the liquid scintillator, achieving a 97.4\% fast neutron rejection power~\cite{Hino:2021uwz}. 

This is also a typical task that can be performed by machine learning algorithms. As an example, a convolutional neural network has been used to discriminate between neutrinoless double-beta decay events and cosmic spallation background in a liquid scintillator detector. The algorithm achieved 61.6\% background rejection at 90\% signal efficiency, without using vertex or energy reconstruction~\cite{Li:2018rzw}.

A Geant4 simulation of the SHiNESS detector shows that a further reduction of a factor of 2 can be achieved by surrounding the steel tank with a layer of 20~cm of paraffin.
Thus, assuming a paraffin shield and a conservative PID rejection factor of 20, the rate of the cosmic-induced neutron background during beam-on periods is $98\pm10$ events/year and in any case can be precisely quantified during beam-off periods. This background must be multiplied by a factor of 2 for the $\nu_e$ CC search, given the extended time window.

\subsubsection*{Accidental backgrounds}\label{sec:acc}

The detection of both the IBD and CC events relies on the detection of a \emph{prompt} signal and a \emph{delayed} coincidence. Several background processes, described below, can mimic at least one of the two. The accidental rate will be given by the sum of the prompt-like background rates multiplied by the sum of the delayed-like background rates, with the same procedure described in ref.~\cite{JSNS2:2013jdh}.

\paragraph{$\bar{\nu}_e$ beam interactions}
The intrinsic $\bar{\nu}_{e}$ beam component, described in section \ref{sec:nue}, can contribute to the accidental background when the prompt or the delayed signal is not detected. This translates into irreducible prompt and delayed accidental rates of $1\times10^{-6}$ for the IBD channel. The IBD positron can also mimic the electron of the CC channel. 

\paragraph{Cosmic rays}
The cosmic-ray background, described in detail in section \ref{sec:cosmic}, can contribute to the accidental background when the prompt recoil proton or the captured neutron is not detected. This corresponds to a small prompt and delayed accidental rate of $3.8\times10^{-7}$ ($7.5\times10^{-7}$) events/spill for IBD (CC) events.

\paragraph{Charged-current $\nu_e$ interactions}
The charged-current channel $^{12}\mathrm{C}(\nu_e, e^-)^{12}\mathrm{N_{gs}}$ of eq. \eqref{eq:cc} can represent a background for the IBD channel search. The $e^-$ emitted promptly can mimic the IBD $e^+$ and the delayed beta decay can fake the neutron absorption. Given the long lifetime of $^{12}\mathrm{N_{gs}}$ (15.8~ms), the probability of detecting both the $e^-$ and the $e^+$ during the beam spill and within the IBD time window (around $100~$\unit{\micro\second}) can be considered negligible. However, they both contribute to the a prompt accidental rate of  $6.4\times10^{-4}$ events/spill and a delayed accidental rate of $0.8\times10^{-4}$ events/spill. The $\nu_e$ can also interact with the carbon atom via $^{12}\mathrm{C}(\nu_e, e^-)^{12}\mathrm{N}^*$, which does not beta decay and contributes only to a prompt accidental rate of $3\times10^{-4}$ events/spill. This background could be eventually suppressed by the topological capabilities of the detector, which permit to distinguish point-like events (such as electrons) from multi-site events (such as photons) in the material~\cite{Wonsak:2019lqq, Dunger:2019dfo}.


\paragraph{Neutron beam component}
Low-energy neutrons (tens of MeV) produced by the beam can be absorbed by the Gd atoms, emitting 8 MeV $\gamma$, mimicking neutron absorption from IBD events. 
The proposal of detecting CE$\nu$NS at the ESS~\cite{Baxter:2019mcx} requires a POT-coincident flux of less than $2\times10^{-3}$ neutrons/cm$^2$ s. With this same requirement, and assuming 0.1\% Gd mass concentration, this flux produces an accidental rate in the detector of $4.5\times10^{-3}$ events/spill after applying preliminary selection cuts. 
However, detailed simulations will be useful to confirm the feasibility of this requirement in the detector location. A dedicated measurement might also be necessary to quantify this component: the JSNS$^2$ collaboration used a 1-ton scintillator detector to precisely assess the neutron rate at J-PARC~\cite{JSNS2:2013jdh}.

\paragraph{Gamma beam component}
Neutrons produced by the beam can be captured in the materials surrounding the detector, producing gamma particles. A fraction of these gammas can eventually enter the detector and mimic a delayed signal in both the IBD and the CC channels. In order to precisely quantify this background, a detailed simulation of the ESS beam and of the experimental hall is necessary. A preliminary result gives a rate of beam-induced gamma of $6.1\times10^{-3}$ events/spill after event selection. 

\medskip

Accidental background rates for $\bar{\nu}_e$ IBD and $\nu_e$ CC channels are summarized in table~\ref{tab:acc}. A conservative cut on the spatial distance between the prompt and the delayed signal can achieve an accidental background rejection power of 90\%, as shown in figure \ref{fig:pos_res_acc}. With this assumption, the accidental background rate is $130\pm17$ events/year for the $\bar{\nu}_e$ IBD channel and $31\pm6$ events/year for the $\nu_e$ CC channel, assuming a 10\% systematic uncertainty.

\begin{table}[htbp]
\centering
\begin{tabular}{l|lll}
\hline
Channel &
Background&
\specialcell{Prompt rate \\(events/spill)} &
\specialcell{Delayed rate\\(events/spill)} \\
\hline
$\bar{\nu}_e$ IBD  & Cosmic rays                                              & $3.8\times10^{-7}$ & $3.8\times10^{-7}$\\
                   & $\bar{\nu}_e$ beam                                       & $1.0\times10^{-6}$ & $1.0\times10^{-6}$\\
                   & $^{12}\mathrm{C}(\nu_e, e^-)^{12}\mathrm{N_{g.s.}}$      & $6.4\times10^{-4}$ & $0.8\times10^{-4}$\\
                   & $^{12}\mathrm{C}(\nu_e, e^-)^{12}\mathrm{N}^*$           & $3.0\times10^{-4}$ & - \\
                   & Beam neutrons                                            & -                  & $4.5\times10^{-3}$\\
                   & Beam gammas                                              & -                  & $6.1\times10^{-3}$\\
\cline{2-4}
                   & Total                                                    & $9.4\times10^{-4}$ & $1.1\times10^{-2}$\\
\hline
$\nu_e$ CC      & Cosmic rays                                                 & $7.5\times10^{-7}$ & $7.5\times10^{-7}$\\
                & $\bar{\nu}_e$ beam                                          & $1.0\times10^{-6}$ & - \\
                & $^{12}\mathrm{C}(\nu_e, e^-)^{12}\mathrm{N}^*$              & $3.0\times10^{-4}$ & - \\
                & Beam gammas                                                 & -                  & $8.1\times10^{-3}$\\
\cline{2-4}
                & Total                                                       & $3.0\times10^{-4}$ & $8.1\times10^{-3}$\\
\hline
\end{tabular}
\caption{\label{tab:acc}%
Rates of prompt and delayed accidental backgrounds for the $\bar{\nu}_e$ IBD and $\nu_e$ CC channels in the SHiNESS detector.}
\end{table}

\subsection{Backgrounds to HNL decay}
\label{sec:hnl-bg}
HNL candidate events are selected by requiring the presence of two leptons originating from the same vertex. In order to obtain an estimate of the background for these events in the experiment, we require the detection of two separate signals in the photosensors within a time window of 100~ns, which makes the accidental background negligible. A primitive lepton-finding reconstruction algorithm has been implemented. The energy of the two signals is required to be above 17~MeV and the angle between the two reconstructed direction must be larger than 15$^{\circ}$. 
Assuming these requirements, the main backgrounds for HNL events are detailed below.

\paragraph{Cosmic-ray background} Cosmic muons can produce radioactive isotopes by spallation on the matter surroundings the detector. These muons do not necessarily pass through the veto, so they can't be tagged. The 17~MeV cut removes the vast majority of these products. The remaining events, misreconstructed as two separate leptons, give a background rate of $18$~events/year for the HNL search.

\paragraph{Charged-current $\nu_e$ interactions} The interaction of a $\nu_e$ with the carbon atom can produce nitrogen atoms in excited states, through the $^{12}\mathrm{C}(\nu_e, e^-)^{12}\mathrm{N}^*$ interaction. Most of these states decay by prompt proton emission to $^{11}$C, which in turn emit gammas between 2~MeV and 7~MeV \cite{LSND:2001fbw}. The 17~MeV cut, coupled with the good energy resolution of the detector, is able to reject essentially all these events and this component is considered negligible. However, the outgoing electron can be misreconstructed as two separate leptons, when the gammas or the delayed beta decay are not detected, giving a background contribution of $14$~events/year. 

\paragraph{Neutral-current $\nu$ interactions} The neutral current interaction between all three neutrino types and carbon atoms, detailed in section \ref{sec:nc}, produce a 15.11~MeV gamma, which could be misreconstructed as two distinct leptons. The vast majority of these events is rejected by the 18~MeV cut and a preliminary simulation gives a background rate for this component of $8$~events/year. 

\paragraph{$\bar{\nu}_e$ beam component} The intrinsic $\bar{\nu}_e$ beam component can represent a background also for the HNL search when the delayed neutron is not detected and the positron is misreconstructed as two distinct leptons. This component gives a background of $21$ events/year.

\vspace{1em}
Given the large uncertainties regarding the reconstruction algorithm and the need for a detailed simulation of the beam gammas and cosmic-ray component, we assign a systematic uncertainty of 50\% to the total HNL background. Thus, the HNL sensitivity calculation of section \ref{sec:res-hnl} assumes a background of $(61\pm31)$~events/year.

\section{\label{sec:results} Results}

This section presents our results for the expected sensitivity of SHiNESS for the three scenarios outlined in section~\ref{sec:pheno}. 

\subsection{\label{sec:res-unitarity} Non-unitarity}

As outlined in section~\ref{sec:unitarity}, SHiNESS is expected to be sensitive to the closure of the unitarity triangle in the $e\mu$ sector, parametrized by $t_{e\mu}$ in eq.~\eqref{eq:unitarity_triang}. In order to determine the attainable sensitivity, we perform a binned analysis of the expected number of IBD interactions. Specifically, we define a Gaussian $\chi^2$ as:
\begin{equation}
\label{eq:chi2_app}
    \chi^2_{\mathrm{IBD}}\left(|t_{e \mu}|^2\right) = \sum_i{\left[\frac{N_{\mathrm{obs,i}}^{\mathrm{IBD}} - N_{\mathrm{exp,i}}^{\mathrm{IBD}}(\xi_\mathrm{sig}, \xi_{\bar\nu_e}, t_{e\mu})}{{\sigma_{\mathrm{stat,i}}}}\right]^2 + \left(\frac{\xi_\mathrm{sig}}{\sigma_{\mathrm{sig}}}\right)^2 + 
    \left(\frac{\xi_{\bar{\nu}_e}}{\sigma_{\bar{\nu}_e}}\right)^2},
\end{equation}
where the sum extends over all energy bins. Here, we assume the experiment will observe a number of events $N_{\mathrm{obs,i}}^\mathrm{IBD} = N_{\mathrm{exp,i}}^\mathrm{IBD} + N_{\mathrm{bkg,i}}$ consistent with the expected IBD interactions from the intrinsic  $\bar{\nu}_e$ beam component, calculated in eq.~\eqref{eq:nue_beam}, plus the steady-state background contribution. In presence of non-unitarity effects, the expected number of events in each bin is obtained as
\begin{equation}
    N_{\mathrm{exp,i}}^{\mathrm{IBD}}=(1 + \xi_\mathrm{sig}) |t_{e \mu}|^2 N^{\bar{\nu}_{\mu}}_{\mathrm{exp,i}} + (1 + \xi_{\bar\nu_e})N_{\mathrm{exp,i}}^\mathrm{IBD} + N_{\mathrm{bkg,i}} \, ,
\end{equation}
where $\xi_{sig}, \xi_{\bar{\nu}_e}$ are the nuisance parameters associated to the signal and the $\bar{\nu}_e$ beam component-normalization, while we assume that uncertainties on the steady-state background are negligible as it can be characterized using beam-off data. This implies that the steady-state background contributions cancel in the numerator of eq.~\eqref{eq:chi2_app} and only contribute to the statistical error in the denominator:  $\sigma_{\mathrm{stat,i}}=\sqrt{N_{\mathrm{exp,i}}^{\mathrm{IBD}} + N_{\mathrm{bkg,i}}}$. Finally, $\sigma_{\mathrm{sys}}=10\%$ is the systematic uncertainty on the normalization, while $\sigma_{\bar{\nu}_e}=25\%$ is the systematic uncertainty on the $n_{\bar{\nu}_e}$ beam component. The minimum of the $\chi^2$ is found after minimizing eq.~\ref{eq:chi2_app} with respect to the nuisance parameters $\xi_{sig}, \xi_{\bar\nu_e}$.

\begin{figure}[htbp]
\centering
\includegraphics[width=0.7\textwidth]{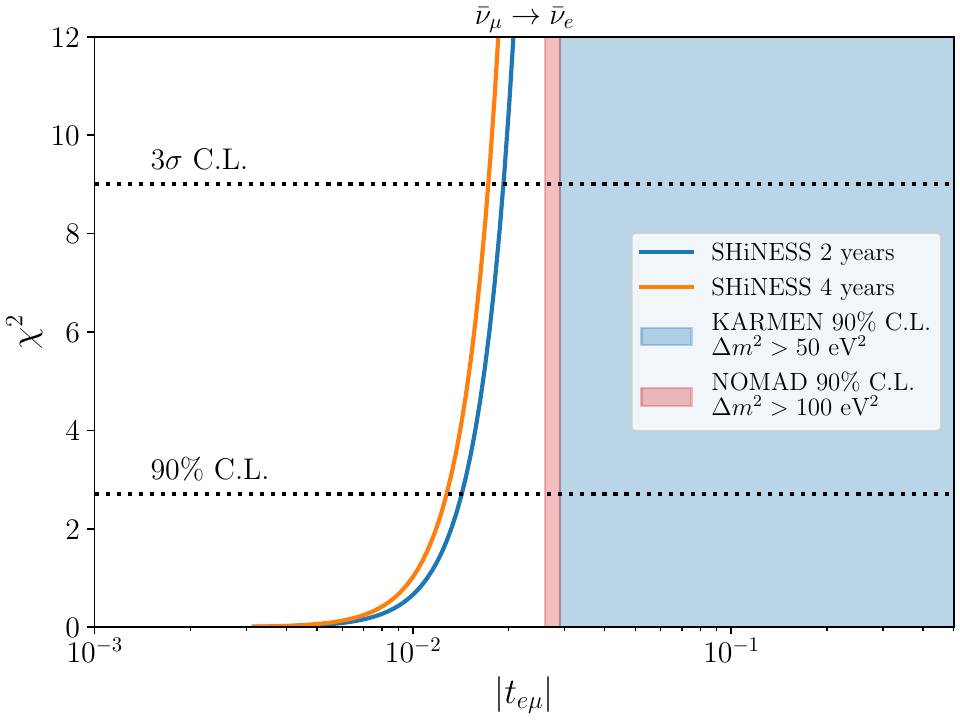}
\caption{$\chi^2$ analysis with one degree of freedom for the $\bar{\nu}_e$ appearance channel as a function of the absolute value of the electron-muon triangle closure parameter $t_{e \mu}$ for 2 years (blue) and 4 years (orange) of data taking. Existing limits at 90\% C.L. from the KARMEN \cite{KARMEN:2002zcm} and NOMAD \cite{NOMAD:2003mqg} experiments are shown in blue and red, respectively.}\label{fig:triangle_emu} 
\end{figure}

Our results are shown in figure~\ref{fig:triangle_emu} for 2 and 4 years of data taking by the blue and orange lines, respectively. As can be seen from the comparison between the blue and orange lines, the improvement obtained by doubling the data taking period is limited by the systematic uncertainties. For comparison, we also show limits on this parameter obtained from the averaged-out region of light sterile neutrino searches at short-baseline neutrino experiments (see the discussion in refs.~\cite{Blennow:2016jkn, Coloma:2021uhq}), using the relation
\begin{equation}
    \left|t_{e \mu}\right|^2 = \frac{1}{2} \sin^2\theta_{\mu e} \, .
\end{equation}
Our results indicate that SHiNESS can significantly improve the limits provided by KARMEN \cite{KARMEN:2002zcm} and NOMAD \cite{NOMAD:2003mqg}, indicated by the shaded regions. 

\subsection{\label{sec:res-steriles} Light sterile neutrinos}

\begin{figure}[htbp]
\begin{tabular}{cc}    
\includegraphics[width=0.48\textwidth]{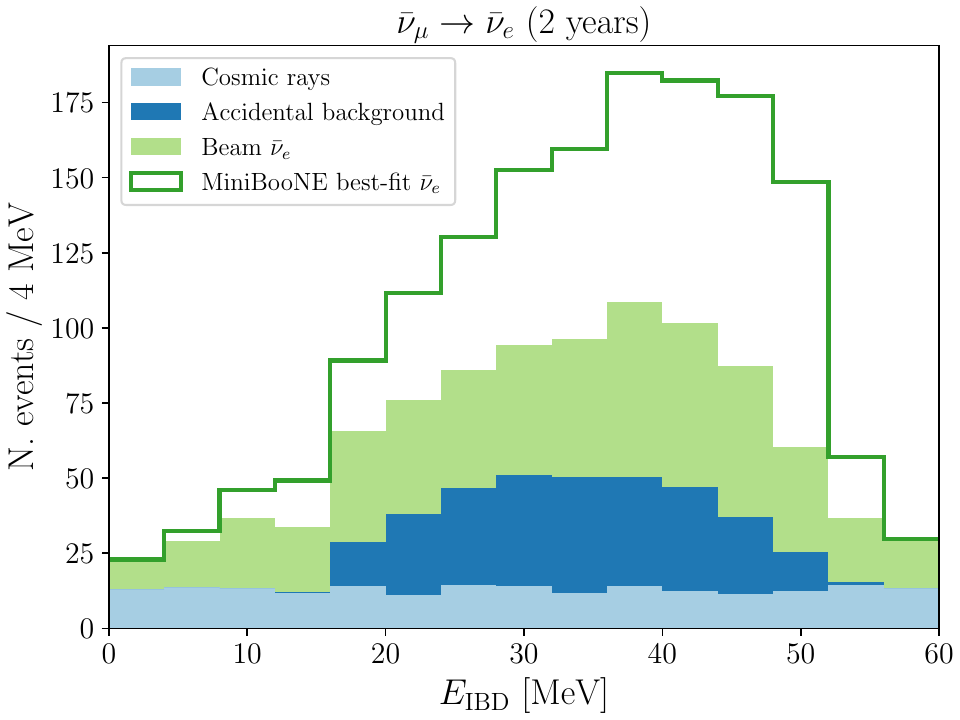} & 
\includegraphics[width=0.48\textwidth]{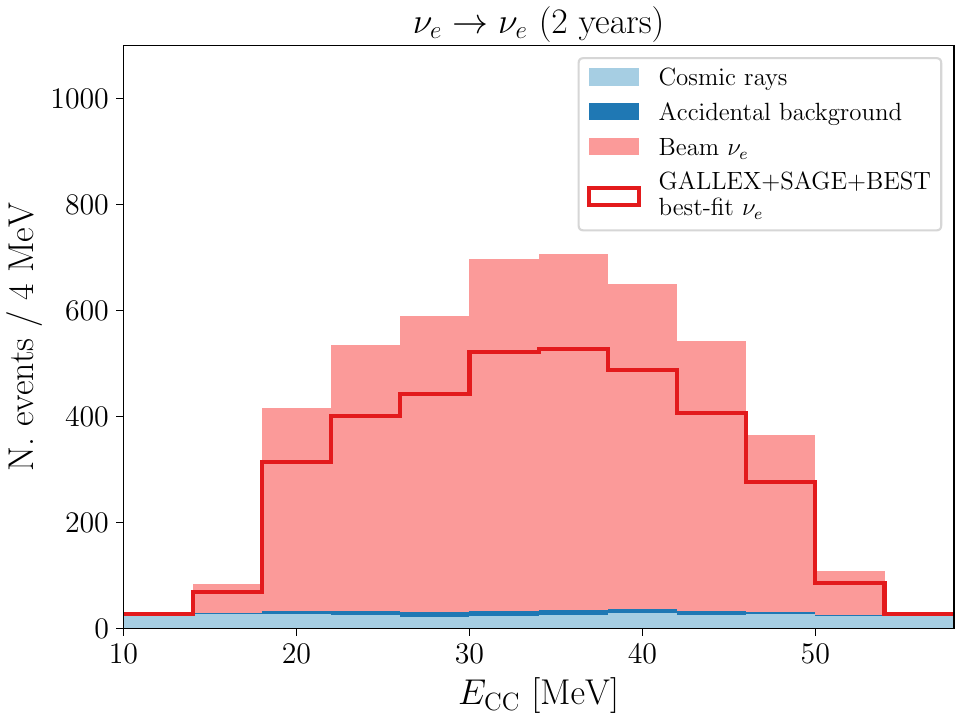}
\end{tabular}
\caption{\label{fig:spectra-osc} \textit{Left:} Reconstructed energy spectrum for IBD candidate events. The filled light blue, blue, and green histograms correspond to the cosmic-ray background, the accidental background, and the $\bar{\nu}_e$ beam component, respectively. The solid green line corresponds to the full spectrum in the presence of $\bar{\nu}_e$ appearance, assuming the MiniBooNE best-fit point $(\sin^22\theta_{\mu e}, \Delta m_{41}^2) = (0.807, 0.043~\mathrm{eV}^2)$~\cite{MiniBooNE:2020pnu}.
\textit{Right:} Reconstructed energy spectrum for $\nu_e$ CC candidate events. The filled light blue, blue, and red histograms correspond to the cosmic-ray background, the accidental background, and the $\nu_e$ beam component in the absence of oscillation, respectively. The solid red line corresponds to the full spectrum in the presence of $\nu_e$ disappearance, assuming the GALLEX+SAGE+BEST best-fit point with Bahcall cross sections \cite{Bahcall:1997eg} $(\sin^22\theta_{ee}, \Delta m_{41}^2) = (0.35, 1.3~\mathrm{eV}^2)$~\cite{Berryman:2021yan}. }
\end{figure}

The search for $\bar{\nu}_{e}$ appearance at spallation sources to look for light sterile neutrinos is not a novel concept: it has been already performed by the KARMEN experiment at the ISIS facility~\cite{KARMEN:2002zcm}, was proposed by the OscSNS collaboration at the SNS~\cite{OscSNS:2013vwh}, and is currently being carried out by the JSNS$^2$ collaboration at J-PARC~\cite{JSNS2:2013jdh}. However, the ESS beam will provide a neutrino flux around 1 order of magnitude larger than J-PARC and SNS, allowing a competitive sensitivity in this channel with a relatively small detector (42~ton of active mass for SHiNESS, compared to the 450~ton of OscSNS). Using eq.~\eqref{eq:Pme} and assuming the MiniBooNE best-fit point at $(\sin^22\theta_{\mu e}, \Delta m_{41}^2) = (0.807, 0.043 ~\mathrm{eV}^2)$~\cite{MiniBooNE:2020pnu} gives a flux-averaged appearance probability of 0.21\% at SHiNESS location. We can multiply this probability by the number of $\bar{\nu}_e$ interactions expected with 100\% $\bar{\nu}_{\mu}\rightarrow\bar{\nu}_e$ transmutation of eq.~\eqref{eq:ibd}, obtaining an appearance signal of $N_{\bar{\nu}_{e}}^{app}=304\pm23$ events/year. Signal and background events for this channel and 2 calendar years of data taking are summarized in the first row of table \ref{tab:summary}. Figure \ref{fig:spectra-osc} shows the reconstructed energy distribution for $\bar{\nu}_e$ candidate events (left panel), assuming a $15\%/\sqrt{E/\mathrm{MeV}}$ energy resolution.

\begin{table}[htbp]
\centering
\begin{tabular}{l|l|ll}
\hline
\specialcell{Channel \\ $(\sin^22\theta, \Delta m_{41}^2)$}&
Signal &
Background &
\\
\hline

\specialcell{$\bar{\nu}_e$ app.\\$(0.807, 0.043 ~\mathrm{eV}^2)$} & $608\pm66$ & $\bar{\nu}_e$ beam & $508\pm129$\\
                                           &            & Cosmic rays        & $196\pm14$\\
                                           &            & Accidental         & $259\pm31$\\

\cline{3-4}
                                           &            & Total              & $963\pm133$\\

\hline
\specialcell{$\nu_e$ disapp.\\$(0.35, 1.3~\mathrm{eV}^2)$} & $1159\pm326$ & Unoscillated $\nu_e$ & $3211\pm326$\\
                                           &            & Cosmic rays        & $372\pm19$\\
                                           &            & Accidental         & $62\pm10$\\
\cline{3-4}
                                           &            & Total              & $3645\pm327$\\

\hline

\end{tabular}
\caption{\label{tab:summary}%
Signal and background events for the $\bar{\nu}_e$ appearance and $\nu_e$ disappearance channels for 2 calendar year of exposure (10000 hours) at full beam power. In the case of $\nu_e$ disappearance, the signal corresponds to the deficit of events. The $(\sin^22\theta, \Delta m_{41}^2)$ values correspond to the MiniBooNE best fit point~\cite{MiniBooNE:2020pnu} for the $\bar{\nu}_e$ appearance and to the GALLEX+SAGE+BEST best fit point~\cite{Berryman:2021yan} with the Bahcall cross sections \cite{Bahcall:1997eg} for the $\nu_e$ disappearance. The uncertainties include a 10\% systematic on the normalization and a 25\% on the $\bar{\nu}_e$ beam component.}
\end{table}

The SHiNESS experiment also has the ability to probe $\nu_e$ disappearance at a short baselines, which is the channel of both the \emph{gallium} and the \emph{reactor antineutrino} anomaly. The best-fit point given by a combined analysis of the BEST, SAGE, and GALLEX results occurs at $(\sin^22\theta_{ee}, \Delta m_{41}^2) = (0.38, 1.3 ~\mathrm{eV}^2)$~\cite{Giunti:2022btk}. Inserting this value into eq.~\eqref{eq:Pee} gives a flux-averaged disappearance probability of 19.1\% at SHiNESS baseline. We can multiply this probability by the number of $\nu_e$ interactions of eq.~\eqref{eq:cc}, obtaining a disappearance signal (thus a deficit of events) of $N_{\nu_{e}}^{disapp}=469 \pm 108$ events/year. Signal and background events for this channel and 2 calendar years of data taking are summarized in the second row of table \ref{tab:summary}. The right panel in figure~\ref{fig:spectra-osc} shows the reconstructed energy distribution for $\nu_e$ candidate events, assuming a  $15\%/\sqrt{E/\mathrm{MeV}}$ energy resolution.

\begin{figure}[htbp]
\begin{tabular}{cc}
\includegraphics[width=0.48\textwidth]{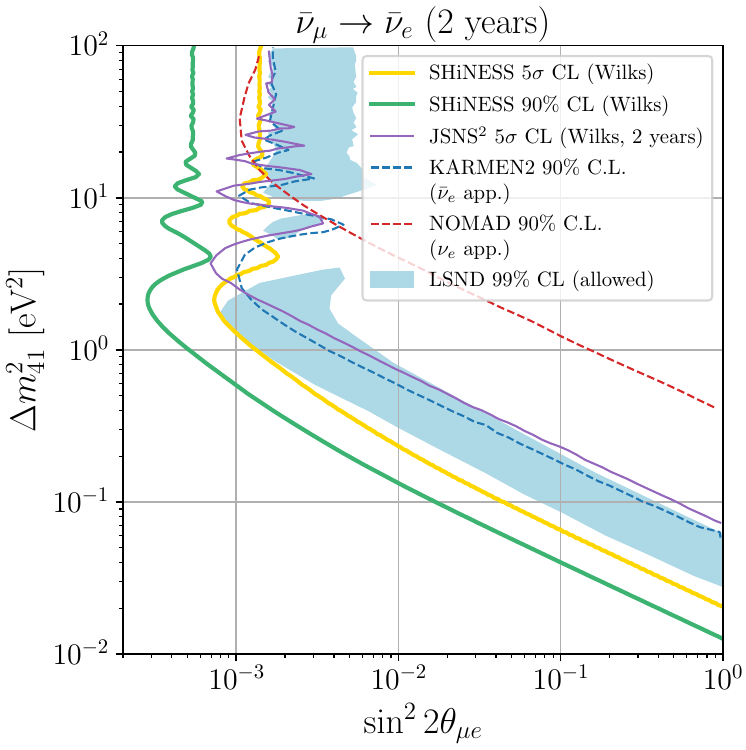} &
\includegraphics[width=0.48\textwidth]{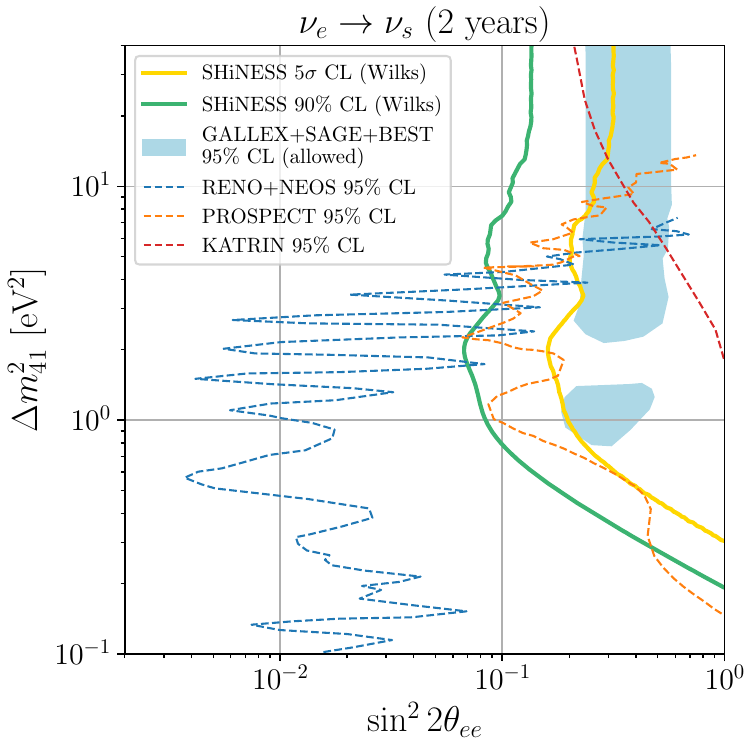}
\end{tabular}
\caption{\label{fig:sens-osc} 
\textit{Left:} The SHiNESS sensitivity to neutrino oscillations in the appearance channel. The LSND allowed region is shown in light blue~\cite{LSND:2001aii}. The thin solid line corresponds to the expected sensitivity of the JSNS$^2$ experiment~\cite{JSNS2:2013jdh}. The dashed lines correspond to exclusion contours of the KARMEN2~\cite{KARMEN:2002zcm} and NOMAD~\cite{NOMAD:2003mqg} experiments.
\textit{Right:} The SHiNESS sensitivity to neutrino oscillations in the $\nu_e$ disappearance channel. The region allowed by the GALLEX+SAGE+BEST combined analysis~\cite{Barinov:2021mjj, Berryman:2021yan} is shown in light blue. The dashed lines correspond to the exclusion contours of the RENO+NEOS~\cite{RENO:2020hva}, PROSPECT~\cite{PROSPECT:2020sxr}, and KATRIN~\cite{KATRIN:2022ith} experiments. Contours labeled as `Wilks' have been obtained assuming the test statistics follows a $\chi^2$ distribution with two degrees of freedom, see text for details. }
\end{figure}

The sensitivity of SHiNESS to light neutrino oscillations is shown in figure~\ref{fig:sens-osc}, computed using a $\chi^2$ analysis analogous to the one of eq.~\eqref{eq:chi2_app}. Results are shown separately for oscillation searches in the appearance (left) and disappearance (right) channels. As shown in the left panel, the SHiNESS experiment is able to fully cover the LNSD 99\% C.L allowed region at a $5\sigma$ level, in a region complementary to the one covered by the JSNS$^2$ experiment~\cite{JSNS2:2013jdh}. As for the search in the disappearance channel, our results indicate that SHiNESS can cover almost entirely the region allowed by a GALLEX+SAGE+BEST combined analysis with a sensitivity better than $5\sigma$, allowing to definitely rule out the (3+1) light sterile neutrino explanation for the gallium anomaly with an independent measurement. Also, as can be seen in both panels, in the limit of large mass-squared splittings the averaged-out oscillations translate into a flat sensitivity. It is easy to check that, once the corresponding value of the mixing angle is mapped onto a unitarity constraint as outlined in refs.~\cite{Blennow:2016jkn, Coloma:2021uhq}, the results in section~\ref{sec:unitarity} are correctly recovered, as expected. 

We finish this section with a remark regarding the validity of the sensitivity limits shown in figure~\ref{fig:sens-osc}. In our calculations, the test statistics is assumed to be distributed according to a $\chi^2$ with 2 degrees of freedom, assuming Wilks' theorem holds. Although this is a common approach in the literature, it has been shown~\cite{Agostini:2019jup, Giunti:2020uhv, PROSPECT:2020raz, Coloma:2020ajw, Berryman:2021yan} that the actual distribution typically differs from a $\chi^2$ distribution, since the conditions needed to apply Wilks' theorem are violated for sterile neutrino oscillation searches. Therefore our results should only be regarded as indicative of an \emph{approximate} value for the attainable sensitivity. An accurate result requires proper evaluation through Monte Carlo simulation: this falls beyond the scope of this article and is left for future work.  

\subsection{\label{sec:res-hnl} Heavy neutral leptons}

\begin{figure}[htbp]
\begin{subfigure}[t]{0.5\textwidth}
\includegraphics[width=1\textwidth]{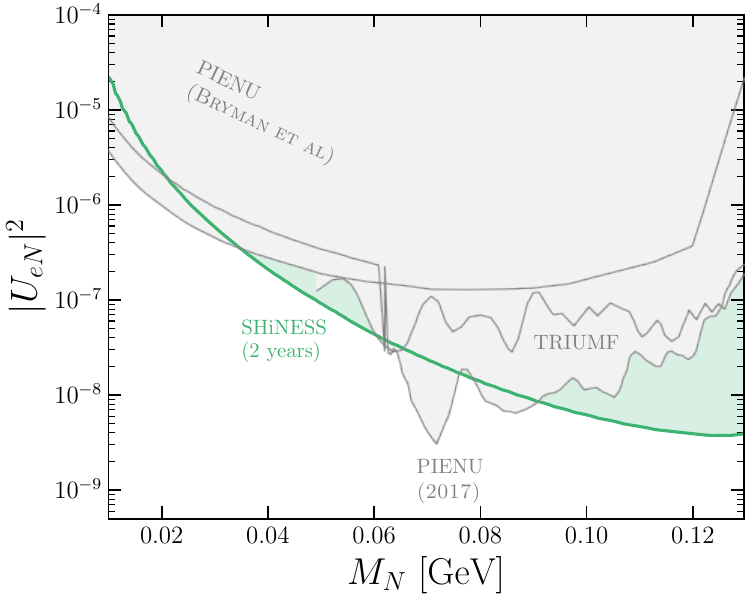}
\caption{\label{fig:emixing} Electron mixing.}
\end{subfigure}
\hfill
\begin{subfigure}[t]{0.5\textwidth}
\includegraphics[width=1\textwidth]{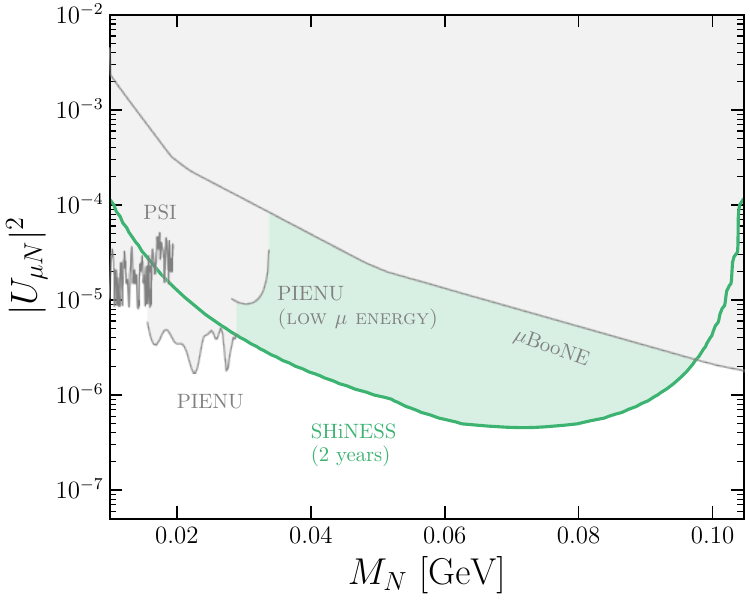}
\caption{\label{fig:mumixing} Muon mixing.}
\end{subfigure}
\caption{\label{fig:hnl} Constraints at 90\% C.L. on the mixing of a HNL $|U_{lN}|^2$ with electron (left) and muon (right) neutrinos, for 2 years of data taking. The gray areas correspond to existing limits from PSI \cite{Daum:1987bg}, PIENU \cite{PIENU:2017wbj, PIENU:2019usb, Bryman:2019bjg}, TRIUMF \cite{Britton:1992xv}, and MicroBooNE \cite{MicroBooNE:2023eef}, obtained from ref.~\cite{Fernandez-Martinez:2023phj}.}
\end{figure}

Figure \ref{fig:hnl} shows the sensitivity at 90\% C.L. for electron and muon mixing in the left and right panels, respectively. In this case, the sensitivity has been computed using just the total rates (unbinned $\chi^2$) for the signal (see section~\ref{sec:hnl-sig}) and backgrounds (see section~\ref{sec:hnl-bg}), assuming a 50\% systematic uncertainty on the background. Within the considered mass range, our results indicate that SHiNESS can set world-leading constraints in both cases, improving over current limits by over an order of magnitude in a sizable region of the parameter space.

\section{Summary and conclusions\label{sec:summary}}

The existence of neutrino masses is irrefutable, and requires the extension of the particle content of the Standard Model (SM). Such extension may lead to additional consequences, including an apparent non-unitary mixing matrix in the active neutrino sector, anomalous oscillations at short-baselines, or even the decay signals of heavy neutral leptons that interact with the visible sector through their mixing with the SM neutrinos.

The ESS provides an exceptional opportunity to carry out neutrino experiments at the edge of the intensity frontier, and to pursue new physics signals. The facility will provide the most intense source of $\pi^+$DAR neutrinos in the world, with one order of magnitude improvement with respect to current facilities. Here we propose to carry out a new experiment, SHiNESS, and highlight the novel opportunities it offers to explore the existence of new physics in the neutrino sector. 

The experiment will be able to collect a high-statistics sample of neutrino interactions with only two calendar years of full-power beam. The baseline detector design includes an acrylic vessel filled with 42~ton of liquid scintillator whose light is detected by large-area PMTs and LAPPDs. Liquid scintillators are a low-risk, proven technology with a solid track record in particle physics. The introduction of LAPPDs allows to separate the Cherenkov light from the scintillation light and enables the reconstruction of topological features. A possible alternative is represented by water-based liquid scintillators, which have the potential to improve the background-rejection capabilities of the experiment. The proposed detector could also operate as a neutrino flux monitor for the proposed CE$\nu$NS detection experiments~\cite{Baxter:2019mcx} and represents a perfect fit for the rich ESS particle physics program~\cite{Abele:2022iml}.

We have identified a potential location for the detector at about 25~m from the beam target (see figure~\ref{fig:plan}). The detector could eventually be replicated at a different baseline in order to further reduce flux systematic uncertainties, as done by the JSNS$^2$ collaboration with the JSNS$^2$-II detector~\cite{Maruyama:2022juu}. 

To illustrate its physics potential, we have computed the expected sensitivities for three well-motivated examples of new physics in the neutrino sector, as outlined above: searches for a non-unitary leptonic mixing matrix (summarized in figure~\ref{fig:triangle_emu}), for anomalous oscillations induced by a sterile neutrino at the eV scale (see figure~\ref{fig:sens-osc}), and decay signals of heavy neutral leptons with masses above the MeV scale (shown in figure~\ref{fig:hnl}). In all three cases considered, we find that the SHiNESS experiment would be able to improve significantly over current constraints, targeting unexplored regions of the parameter space. A notebook has also been made available online, which allows to reproduce the sensitivities and energy spectra of the channels used for our oscillation analysis~\cite{binder}.

The timing of this proposal is timely and appropriate, as the ESS user program is expected to start in 2025. The search for light sterile neutrinos has spurred a very active physics program in several laboratories worldwide. At the same time, the renewed interest in heavy neutral leptons at the MeV-GeV scale has triggered an intense activity in the field. In this sense we underscore the short timescale of the proposed experiment, which could deliver results in a very short period of time. To begin with, the SHiNESS detector could start acquiring data as soon as the beam turns on, and does not require any beamline upgrade or modification. Moreover, thanks to the intensity of the ESS beam, SHiNESS has the capability to reach competitive sensitivities for several new physics scenarios in a couple of years of data taking.

\acknowledgments

The authors are thankful to R. Castillo Fernandez, P. Ferrario, P. Huber, J. Martín-Albo, F. Monrabal, P. Novella, M. Ovchynnikov, M. Sorel, and D. Zerzion for the precious suggestions.

SRS acknowledges the support of a fellowship from ``la Caixa Foundation'' (ID 100010434) with code LCF/BQ/PI22/11910019 and of the Severo Ochoa Program grant CEX2018-000867-S. This work has received partial support from the European Union’s Horizon 2020 Research and Innovation Programme under the Marie Sklodowska-Curie grant agreement no.\ 860881-HIDDeN and the Marie Sklodowska-Curie Staff Exchange grant agreement no.\ 101086085–ASYMMETRY. 
PC acknowledges financial support by grant RYC2018-024240-I, funded by MCIN/AEI/10.13039/501100011033 and by ``ESF Investing in your future''. She is also supported by the Spanish Research Agency (Agencia Estatal de Investigaci\'on) through the grant IFT Centro de Excelencia Severo Ochoa no.\ CEX2020-001007-S funded by MCIN/AEI/10.13039/501100011033/, and by Grant PID2022-142545NB-C21, funded by MCIN/AEI/10.13039/501100011033/ FEDER, UE.


\bibliographystyle{JHEP}
\bibliography{biblio.bib}


\end{document}